\documentclass[twocolumn]{aastex631}

\shorttitle{Measuring LyC \fesc from \uvc}
\shortauthors{Wang et al. (2025)}

\newcommand{\Ntot}{61\xspace}
\newcommand{\Nstack}{56\xspace}
\newcommand{\Ndet}{5\xspace}
\newcommand{\lowz}{\textsc{low-\ensuremath{z}}\xspace}
\newcommand{\medz}{\textsc{intermediate-\ensuremath{z}}\xspace}

\newcommand{\goldz}{\textsc{gold-\ensuremath{z}}\xspace}
\newcommand{\snOII}{\textsc{strong-\OII}\xspace}
\newcommand{\hiMuv}{\textsc{faint-\ensuremath{M_{\rm UV}}}\xspace}
\newcommand{\lowEbv}{\textsc{low-E(B-V)}\xspace}
\newcommand{\Flowz}{28\xspace}
\newcommand{\Fmedz}{28\xspace}

\newcommand{\Glowz}{12\xspace}
\newcommand{\Gmedz}{7\xspace}

\newcommand{\Olowz}{11\xspace}
\newcommand{\Omedz}{15\xspace}

\newcommand{\tao}{\textsc{TAOIST\_MC}\xspace}

\usepackage[english]{babel} 
\usepackage[utf8]{inputenc} 
\usepackage[T1]{fontenc}    
\usepackage{ae,aecompl}
\usepackage{pgf,pgfarrows,pgfnodes,pgfautomata,pgfheaps}
\usepackage{graphicx}       
\usepackage{natbib}         
\usepackage{url}            
\usepackage{grffile}        
\usepackage{mathtools}      
\usepackage{multirow}       
\usepackage{xspace}         

\usepackage{tikz,lipsum,lmodern}
\usepackage[most]{tcolorbox}

\usepackage{amsmath,amssymb,amsxtra,amsfonts}   
\usepackage{txfonts}

\usepackage{color}         
\definecolor{gold}{rgb}{1,0.80,0}
\definecolor{orange}{rgb}{1,0.5,0}
\definecolor{midgray}{gray}{0.3}
\definecolor{lblue}{rgb}{0,0.2,0.6}
\definecolor{dgreen}{rgb}{0.1,0.6,0.3}
\definecolor{purple}{rgb}{0.5019607843137255,0.0,0.5019607843137255}

\renewcommand\farcs{\mbox{$.\!^{\prime\prime}$}}    
\renewcommand{\arcsec}{\mbox{$^{\prime\prime}$}\xspace}  
\renewcommand{\arcdeg}{\mbox{$^{\circ}$}\xspace}             

\newcommand{\be}{\begin{equation}}
\newcommand{\ee}{\end{equation}}

\newcommand{\ba}{\begin{align}}
\newcommand{\ea}{\end{align}}

\DeclareMathOperator{\ud}{d}

\newcommand{\Msun}{\ensuremath{M_\odot}\xspace}

\newcommand{\Mstar}{\ensuremath{M_\ast}\xspace}
\newcommand{\Lstar}{\ensuremath{L_\ast}\xspace}

\newcommand{\fesc}{\ensuremath{f_{\rm esc}}\xspace}

\newcommand{\tauIGM}{\ensuremath{\tau_{\rm IGM}}\xspace}

\newcommand{\K}{\ensuremath{\rm K}\xspace}

\newcommand{\Funit}{\ensuremath{\rm erg~s^{-1}~cm^{-2}}\xspace}

\newcommand{\Ha}{\textrm{H}\ensuremath{\alpha}\xspace}
\newcommand{\Hb}{\textrm{H}\ensuremath{\beta}\xspace}

\newcommand{\HI}{\textrm{H}\textsc{i}\xspace}

\newcommand{\OII}{[\textrm{O}~\textsc{ii}]\xspace}
\newcommand{\OIII}{[\textrm{O}~\textsc{iii}]\xspace}

\def\U{\ensuremath{U_{360}}\xspace}     
\def\B{\ensuremath{B_{435}}\xspace}
\def\V{\ensuremath{V_{606}}\xspace}
\def\I{\ensuremath{I_{814}}\xspace}
\def\Y{\ensuremath{Y_{105}}\xspace}
\def\J{\ensuremath{J_{125}}\xspace}

\def\H{\ensuremath{H_{160}}\xspace}

\newcommand{\sex}{\textsc{SExtractor}\xspace}

\newcommand{\adriz}{\textsc{AstroDrizzle}\xspace}

\newcommand{\grzl}{\textsc{Grizli}\xspace}

\newcommand{\jwst}{\textit{JWST}\xspace}

\newcommand{\hst}{\textit{HST}\xspace}

\newcommand{\candels}{\textit{CANDELS}\xspace}
\newcommand{\uvc}{\textit{UVCANDELS}\xspace}

\newcommand{\mosfire}{\textit{MOSFIRE}\xspace}

\def\ie{i.e.\xspace}
\def\eg{e.g.\xspace}
\def\etc{etc.\xspace}

\def\p{{\rm prior}}

\newcommand{\n}{\ensuremath{{\nu}\rm}\xspace}

\usepackage{etoolbox}
\makeatletter
\patchcmd{\NAT@citex}
  {\@citea\NAT@hyper@{\NAT@nmfmt{\NAT@nm}\NAT@date}}
  {\@citea\NAT@nmfmt{\NAT@nm}\NAT@hyper@{\NAT@date}}
  {}
  {}
\patchcmd{\NAT@citex}
  {\@citea\NAT@hyper@{%
     \NAT@nmfmt{\NAT@nm}%
     \hyper@natlinkbreak{\NAT@aysep\NAT@spacechar}{\@citeb\@extra@b@citeb}%
     \NAT@date}}
  {\@citea\NAT@nmfmt{\NAT@nm}%
   \NAT@aysep\NAT@spacechar%
   \NAT@hyper@{\NAT@date}}
  {}
  {}
\patchcmd{\NAT@citex}
  {\@citea\NAT@hyper@{%
     \NAT@nmfmt{\NAT@nm}%
     \hyper@natlinkbreak{\NAT@spacechar\NAT@@open\if*#1*\else#1\NAT@spacechar\fi}%
       {\@citeb\@extra@b@citeb}%
     \NAT@date}}
  {\@citea\NAT@nmfmt{\NAT@nm}%
   \NAT@spacechar\NAT@@open\if*#1*\else#1\NAT@spacechar\fi%
   \NAT@hyper@{\NAT@date}}
  {}
  {}
\makeatother

\graphicspath{{./}{figures/}}
\begin{document}


\title{The Lyman Continuum Escape Fraction of Star-forming Galaxies at $2.4\lesssim z\lesssim3.0$ from \uvc}

\correspondingauthor{Xin Wang}
\email{xwang@ucas.ac.cn}

\author[0000-0002-9373-3865]{Xin Wang}
\affil{School of Astronomy and Space Science, University of Chinese Academy of Sciences (UCAS), Beijing 100049, China}
\affil{National Astronomical Observatories, Chinese Academy of Sciences, Beijing 100101, China}
\affil{Institute for Frontiers in Astronomy and Astrophysics, Beijing Normal University,  Beijing 102206, China}

\author[0000-0002-7064-5424]{Harry I. Teplitz}
\affil{IPAC, Mail Code 314-6, California Institute of Technology, 1200 E. California Blvd., Pasadena CA, 91125, USA}

\author[0000-0002-0648-1699]{Brent M. Smith}
\affil{School of Earth and Space Exploration, Arizona State University, Tempe, AZ 85287, USA}

\author[0000-0001-8156-6281]{Rogier A. Windhorst} 
\affiliation{School of Earth and Space Exploration, Arizona State University, Tempe, AZ 85287, USA}

\author[0000-0002-9946-4731]{Marc Rafelski}
\affiliation{Space Telescope Science Institute, Baltimore, MD 21218, USA}
\affiliation{Department of Physics and Astronomy, Johns Hopkins University, Baltimore, MD 21218, USA}

\author[0000-0001-7166-6035]{Vihang Mehta}
\affiliation{IPAC, Mail Code 314-6, California Institute of Technology, 1200 E. California Blvd., Pasadena CA, 91125, USA}

\author[0000-0002-8630-6435]{Anahita Alavi}
\affiliation{IPAC, Mail Code 314-6, California Institute of Technology, 1200 E. California Blvd., Pasadena CA, 91125, USA}

\author[0000-0001-7673-2257]{Zhiyuan Ji}
\affiliation{Department of Astronomy, University of Massachusetts, Amherst, MA 01003, USA}

\author[0000-0003-2680-005X]{Gabriel Brammer}
\affiliation{Cosmic Dawn Center (DAWN), Denmark}
\affiliation{Niels Bohr Institute, University of Copenhagen, Jagtvej 128, DK-2200 Copenhagen N, Denmark}

\author[0000-0001-6482-3020]{James Colbert}
\affiliation{IPAC, Mail Code 314-6, California Institute of Technology, 1200 E. California Blvd., Pasadena CA, 91125, USA}

\author[0000-0001-9440-8872]{Norman Grogin}
\affiliation{Space Telescope Science Institute, Baltimore, MD 21218, USA}

\author[0000-0001-6145-5090]{Nimish P. Hathi}
\affiliation{Space Telescope Science Institute, Baltimore, MD 21218, USA}

\author[0000-0002-6610-2048]{Anton M. Koekemoer}
\affiliation{Space Telescope Science Institute, Baltimore, MD 21218, USA}

\author[0000-0002-0604-654X]{Laura Prichard}
\affiliation{Space Telescope Science Institute, Baltimore, MD 21218, USA}

\author[0000-0002-9136-8876]{Claudia Scarlata}
\affiliation{Minnesota Institute of Astrophysics and School of Physics and Astronomy, University of Minnesota, Minneapolis, MN 55455, USA}

\author[0000-0003-3759-8707]{Ben Sunnquist}
\affiliation{Space Telescope Science Institute, Baltimore, MD 21218, USA}

\author[0000-0002-7959-8783]{Pablo Arrabal Haro}
\affiliation{NSF's NOIRLab, Tucson, AZ 85719, USA}

\author[0000-0003-1949-7638]{Christopher Conselice}
\affiliation{School of Physics and Astronomy, The University of Nottingham, University Park, Nottingham NG7 2RD, UK}

\author[0000-0003-1530-8713]{Eric Gawiser}
\affiliation{Department of Physics and Astronomy, Rutgers, The State University of New Jersey, Piscataway, NJ 08854, USA}

\author[0000-0003-2775-2002]{Yicheng Guo}
\affiliation{Department of Physics and Astronomy, University of Missouri, Columbia, MO 65211, USA}

\author[0000-0001-8587-218X]{Matthew Hayes}
\affiliation{Stockholm University, Department of Astronomy and Oskar Klein Centre for Cosmoparticle Physics, AlbaNova University Centre, SE-10691, Stockholm, Sweden}

\author[0000-0003-1268-5230]{Rolf A. Jansen}
\affiliation{School of Earth and Space Exploration, Arizona State University, Tempe, AZ 85287, USA}

\author[0000-0003-1581-7825]{Ray A. Lucas}
\affiliation{Space Telescope Science Institute, Baltimore, MD 21218, USA}

\author[0000-0002-8190-7573]{Robert O'Connell}
\affiliation{Department of Astronomy, University of Virginia, Charlottesville, VA 22904}

\author[0000-0002-4271-0364]{Brant Robertson}
\affiliation{Department of Astronomy and Astrophysics, University of California, Santa Cruz, Santa Cruz, CA 95064, USA}

\author[0000-0003-3527-1428]{Michael Rutkowski}
\affiliation{Department of Physics and Astronomy, Minnesota State University Mankato, Mankato, MN 56001, USA}

\author[0000-0002-4935-9511]{Brian Siana}
\affiliation{Department of Physics and Astronomy, University of California, Riverside, Riverside, CA 92521, USA}

\author[0000-0002-5057-135X]{Eros Vanzella}
\affiliation{INAF -- OAS, Osservatorio di Astrofisica e Scienza dello Spazio di Bologna, via Gobetti 93/3, I-40129 Bologna, Italy}


\author[0000-0003-4439-6003]{Teresa Ashcraft}
\affiliation{School of Earth and Space Exploration, Arizona State University, Tempe, AZ 85287, USA}
\author[0000-0002-9921-9218]{Micaela Bagley}
\affiliation{Department of Astronomy, The University of Texas at Austin, Austin, TX 78712, USA}
\author[0000-0003-0556-2929]{Ivano Baronchelli}
\affiliation{INAF-Osservatorio Astronomico di Padova, Vicolo dell'Osservatorio 5, I-35122, Padova, Italy}
\affiliation{Dipartimento di Fisica e Astronomia, Università di Padova, vicolo Osservatorio, 3, I-35122 Padova, Italy}
\author[0000-0001-6813-875X]{Guillermo Barro}
\affiliation{Department of Physics, University of the Pacific, Stockton, CA 95211, USA}
\author[0000-0003-2102-3933]{Alex Blanche}
\affiliation{School of Earth and Space Exploration, Arizona State University, Tempe, AZ 85287, USA}
\author[0000-0002-7767-5044]{Adam Broussard}
\affiliation{Department of Physics and Astronomy, Rutgers, The State University of New Jersey, Piscataway, NJ 08854, USA}
\author[0000-0001-6650-2853]{Timothy Carleton}
\affiliation{School of Earth and Space Exploration, Arizona State University, Tempe, AZ 85287, USA}
\author[0000-0003-3691-937X]{Nima Chartab}
\affiliation{Observatories of the Carnegie Institution of Washington, Pasadena, CA 91101, US}
\author[0000-0001-8551-071X]{Yingjie Cheng}
\affiliation{Department of Astronomy, University of Massachusetts, Amherst, 01003 MA, USA}
\author[0000-0001-9560-9174]{Alex Codoreanu}
\affiliation{Centre for Astrophysics and Supercomputing, Swinburne University of Technology, Hawthorn VIC 3122, Australia}
\author[0000-0003-3329-1337]{Seth Cohen} 
\affiliation{School of Earth and Space Exploration, Arizona State University, Tempe, AZ 85287, USA}
\author[0000-0002-7928-416X]{Y. Sophia Dai}
\affiliation{National Astronomical Observatories, Chinese Academy of Sciences, Beijing 100101, China}
\author[0000-0003-4919-9017]{Behnam Darvish}
\affiliation{Department of Physics and Astronomy, University of California, Riverside, Riverside, CA 92521, USA}
\author[0000-0003-2842-9434]{Romeel Dav\'{e}}
\affiliation{Institute for Astronomy, University of Edinburgh, Edinburgh, EH9 3HJ, UK}
\author[0000-0001-9022-665X]{Laura DeGroot}
\affiliation{College of Wooster, Wooster, OH 44691, USA}
\author{Duilia De Mello}
\affiliation{Department of Physics, The Catholic University of America, Washington, DC 20064, USA}
\author[0000-0001-5414-5131]{Mark Dickinson}
\affiliation{NSF's NOIRLab, Tucson, AZ 85719, USA}
\author[0000-0003-2047-1689]{Najmeh Emami}
\affiliation{Department of Physics and Astronomy, University of California, Riverside, Riverside, CA 92521, USA}
\author[0000-0001-7113-2738]{Henry Ferguson}
\affiliation{Space Telescope Science Institute, Baltimore, MD 21218, USA}
\author[0000-0002-8919-079X]{Leonardo Ferreira}
\affiliation{Centre for Astronomy and Particle Physics, School of Physics and Astronomy, University of Nottingham, NG7 2RD, UK}
\author[0000-0003-0792-5877]{Keely Finkelstein}
\affiliation{Department of Astronomy, The University of Texas at Austin, Austin, TX 78712, USA}
\author[0000-0001-8519-1130]{Steven Finkelstein}
\affiliation{Department of Astronomy, The University of Texas at Austin, Austin, TX 78712, USA}
\author[0000-0003-2098-9568]{Jonathan P. Gardner}
\affiliation{Astrophysics Science Division, NASA Goddard Space Flight Center, Greenbelt, MD 20771, USA}
\author[0000-0002-7732-9205]{Timothy Gburek}
\affiliation{Department of Physics and Astronomy, University of California, Riverside, Riverside, CA 92521, USA}
\author[0000-0002-7831-8751]{Mauro Giavalisco}
\affiliation{Department of Astronomy, University of Massachusetts, Amherst, MA 01003, USA}
\author[0000-0002-5688-0663]{Andrea Grazian}
\affiliation{INAF-Osservatorio Astronomico di Padova, Vicolo dell'Osservatorio 5, I-35122, Padova, Italy}
\author[0000-0001-6842-2371]{Caryl Gronwall}
\affiliation{Department of Astronomy \& Astrophysics, The Pennsylvania State University, University Park, PA 16802, USA}
\affiliation{Institute for Gravitation and the Cosmos, The Pennsylvania State University, University Park, PA 16802, USA}
\author[0000-0003-2226-5395]{Shoubaneh Hemmati}
\affiliation{IPAC, Mail Code 314-6, California Institute of Technology, 1200 E. California Blvd., Pasadena CA, 91125, USA}
\author[0000-0002-5924-0629]{Justin Howell}
\affiliation{IPAC, Mail Code 314-6, California Institute of Technology, 1200 E. California Blvd., Pasadena CA, 91125, USA}
\author[0000-0001-9298-3523]{Kartheik Iyer}
\affiliation{Dunlap Institute for Astronomy \& Astrophysics, University of Toronto, 50 St George Street, Toronto, ON M5S 3H4, CA}
\author[0000-0002-5601-575X]{Sugata Kaviraj}
\affiliation{Centre for Astrophysics Research, Department of Physics, Astronomy and Mathematics, University of Hertfordshire, Hatfield, AL10 9AB, UK}
\author[0000-0002-8816-5146]{Peter Kurczynski}
\affiliation{Astrophysics Science Division, NASA Goddard Space Flight Center, Greenbelt, MD 20771, USA}
\author[0009-0000-1797-0300]{Ilin Lazar}
\affiliation{Department of Galaxies and Cosmology, Max Planck Institute for Astronomy, K\"{o}nigstuhl 17, 69117 Heidelberg}
\author[0000-0001-6529-8416]{John MacKenty}
\affiliation{Space Telescope Science Institute, Baltimore, MD 21218, USA}
\author[0000-0002-6016-300X]{Kameswara Bharadwaj Mantha}
\affiliation{Minnesota Institute of Astrophysics and School of Physics and Astronomy, University of Minnesota, Minneapolis, MN 55455, USA}
\author[0000-0002-6632-4046]{Alec Martin}
\affiliation{Department of Physics and Astronomy, University of Missouri, Columbia, MO 65211, USA}
\author[0000-0003-2939-8668]{Garreth Martin}
\affiliation{Korea Astronomy and Space Science Institute, Yuseong-gu, Daejeon 34055, Korea}
\affiliation{Steward Observatory, University of Arizona, Tucson, AZ 85719, USA}
\author[0000-0002-5506-3880]{Tyler McCabe}
\affiliation{School of Earth and Space Exploration, Arizona State University, Tempe, AZ 85287, USA}
\author[0000-0001-5846-4404]{Bahram Mobasher}
\affiliation{Department of Physics and Astronomy, University of California, Riverside, Riverside, CA 92521, USA}
\author[0000-0001-5294-8002]{Kalina Nedkova}
\affiliation{Department of Physics and Astronomy, Johns Hopkins University, Baltimore, MD 21218, USA}
\author[0000-0002-8085-7578]{Charlotte Olsen}
\affiliation{Department of Physics and Astronomy, Rutgers, The State University of New Jersey, Piscataway, NJ 08854, USA}
\author[0000-0001-9665-3003]{Lillian Otteson}
\affiliation{School of Earth and Space Exploration, Arizona State University, Tempe, AZ 85287, USA}
\author[0000-0002-5269-6527]{Swara Ravindranath}
\affiliation{Space Telescope Science Institute, Baltimore, MD 21218, USA}
\author[0000-0002-9961-2984]{Caleb Redshaw}
\affiliation{School of Earth and Space Exploration, Arizona State University, Tempe, AZ 85287, USA}
\author[0000-0002-0364-1159]{Zahra Sattari}
\affiliation{Department of Physics and Astronomy, University of California, Riverside, Riverside, CA 92521, USA}
\author[0000-0002-2390-0584]{Emmaris Soto}
\affiliation{Computational Physics, Inc., Springfield, VA 22151, USA}
\author[0000-0003-3466-035X]{L. Y. Aaron Yung}
\affiliation{Astrophysics Science Division, NASA Goddard Space Flight Center, Greenbelt, MD 20771, USA}
\author[0000-0002-7830-363X]{Bonnabelle Zabelle}
\affiliation{Minnesota Institute of Astrophysics and School of Physics and Astronomy, University of Minnesota, Minneapolis, MN 55455, USA}
\author{the UVCANDELS team}


\begin{abstract}
    The UltraViolet Imaging of the Cosmic Assembly Near-infrared Deep Extragalactic Legacy Survey Fields (\uvc) survey is a Hubble Space Telescope (\hst) Cycle-26 Treasury Program, allocated in total 164 orbits of primary Wide-Field Camera 3 (WFC3) Ultraviolet and VISible light (UVIS) F275W imaging with coordinated parallel Advanced Camera for Surveys (ACS) F435W imaging, on four of the five premier extragalactic survey fields: GOODS-N, GOODS-S, EGS, and COSMOS.
    We introduce this survey by presenting a comprehensive analysis of the absolute escape fraction ($\fesc^{\rm abs}$) of Lyman continuum (LyC) radiation through stacking the UV images of a population of star-forming galaxies with secure redshifts at $2.4\leq z\leq3.0$.
    Our stacking benefits from the catalogs of high-quality spectroscopic redshifts compiled from archival ground-based data and \hst slitless spectroscopy, carefully vetted by dedicated visual inspection efforts.
    We develop a robust stacking method to apply to 10 samples of in total 56 galaxies,
    and perform detailed Monte Carlo (MC) simulations of the intergalactic medium (IGM) attenuation, to take into account the sample variance of the mean IGM transmission when measuring $\fesc^{\rm abs}$.
    The full stack at $z\approx2.44$ from 28 galaxies places a stringent 1-$\sigma$ upper limit of $\fesc^{\rm abs}\lesssim5\%$, whereas the full stack at $z\approx2.72$ of equal number of galaxies gives an upper limit of $\fesc^{\rm abs}\lesssim26\%$ at 1-$\sigma$ confidence level.
    These new F275W and F435W imaging mosaics from \uvc have been made publicly available on the Barbara A. Mikulski Archive for Space Telescopes (MAST).
\end{abstract}


\section{Introduction} \label{sect:intro}

Cosmic reionization is the last major phase transition of the Universe when its bulk properties are altered by galaxies \cite[see \eg,][]{Fan.2006,Stark.2016}. There has been increasing evidence that young massive stars collectively dominate over supermassive black holes in the early Universe in producing the LyC photons (with rest-frame wavelength $\lambda_{\rm rest}$<912\AA) that reionize the neutral IGM at $z$\,$\sim$6-9 \citep{Dayal.2020}.
However the exact fraction of these LyC photons that evade photoelectric and dust absorption from their origin galaxy and escape into the IGM to ionize it --- \fesc --- remains elusive and controversial \citep{Finkelstein.2019,Naidu.2020}, making \fesc one of the greatest unknowns in reionization studies.
Furthermore, the intervening IGM transmission drops precipitously at $z\geq4$, due to the steeply rising IGM opacity, \tauIGM \citep{Inoue.2014}.
As a result, it is basically impractical to directly measure any LyC leakage in the epoch of reionization (EoR).

A more feasible way forward is thus to find low-redshift galaxies that are analogous to those $\gtrsim$13 Gyrs ago, which are thought to drive the reionization process. 
The constraints and measurements of their LyC \fesc and its correlation with their physical properties can help shed light upon the detailed physical mechanism conducive to the escape of the ionizing radiation into the IGM, surviving the absorption by the intervening \HI gas and dust in the interstellar medium (ISM) or circumgalactic medium (CGM) \citep{Steidel.2018}.

Due to the intrinsic faintness of LyC flux, the searches for LyC leakage at any cosmological distances have been very challenging.
Generally speaking, currently available ultraviolet (UV) instrumentation leads to fruitful detection of LyC leakage at primarily two redshift windows. 
One is at $z\sim0.3$, where LyC can be captured by the space-based far-UV spectroscopy \citep[see \eg,][]{Leitherer.1995,Bergvall.2006,Borthakur.2014,Izotov.2016zyg,Izotov.201878g}.
The recent \hst Cosmic Origin Spectrograph (COS) campaigns have led to the detection of LyC signals in several dozen targets at $z\sim0.3$, which are usually compact isolated galaxies with high ionization and extreme star formation, a.k.a. Green Peas (GPs) \citep[see \eg,][]{Izotov.2021,Flury.2022}.
Yet the complex selection function and dissimilar environments of these $z\sim0.3$ GPs make it difficult to generalize these findings to the EoR \citep{Naidu.2021}.

The other major redshift range, much closer to the EoR in time, is at $z$\,$\sim$2-3 where LyC flux is accessible to ground-based {blue-sensitive instruments} \citep[see \eg][]{Marchi.2018,Steidel.2018,Pahl.2021,Saxena.2021} or \hst WFC3 UVIS imaging \citep[see \eg][]{Vanzella.2016,Bian.2017,Naidu.2017,Fletcher.2019,Rivera-Thorsen.2019,Yuan.2021,Prichard.2022}. 
These studies are progressively reaching a consensus in support of an intriguing bimodality of \fesc.
{
On one hand, there exists a population of sources with super-\Lstar UV luminosities that leak surprisingly large amount of ionizing radiation with $\fesc\gtrsim50\%$, \eg, 
\textsc{Ion1} at $z=3.8$ \citep{Vanzella.2012,Ji.2020}, 
\textsc{Ion2} at $z=3.2$ \citep{Barros.2016,Vanzella.2016},
\textsc{Ion3} at $z=4$ \citep{Vanzella.2018}, 
the \textsc{Sunburst} arc at $z=2.37$ \citep{Rivera-Thorsen.2019}, 
\textsc{Q1549-C25} at $z=3.2$ \citep{,Shapley.2016}, \etc.
}
On the other hand, there are far more galaxies on which only upper limits can be derived, despite deep imaging and spectroscopy (tens of hours of integration on a 8/10-m diameter ground-based telescope or \hst) \citep{Malkan.2003,Siana.2007,Siana.2010,Siana.2015,Vasei.2016,Alavi.2020}.
Stacking these non-detections produces population-averaged \fesc with tight upper limits of 5-10\% for $\gtrsim$0.5$L_\ast$ galaxies at $z\sim3$ \citep{Rutkowski.2017,Steidel.2018,Begley.2022}.
At face value, this tight upper limit does not bode well for the completion of reionization by $z\sim6$, which requires \fesc$\sim$10-20\% under the canonical reionization picture \citep{Robertson.2013,Robertson.2015}.
This crisis can be largely alleviated though, if one or several of the following conditions are met.
1) Galaxies produce and leak ionzing photons more efficiently at higher redshifts and lower stellar mass \citep{Finkelstein.2019,masciaClosingSourcesCosmic2023,atek.2024}.
2) Reionization is instead driven by a much rarer class of massive, bright galaxies ($M_{\rm UV} \lesssim -20$ and $\Mstar\gtrsim10^{8.5}\Msun$) with much higher \fesc and ionizing photon production efficiency \citep[the oligarch scenario,][]{Naidu.2020,Naidu.2021}.
3) The completion of reionization is delayed to redshift of $z\sim5$ \citep{Kulkarni.2019iz8}.
Furthermore, \fesc is closely tied to other galaxy properties (stellar mass, star formation rate, stellar initial mass function, metallicity, \etc), while the overall emissivity caused by the entire galaxy population also depends on the shape and cutoff of the luminosity function.

The Ultraviolet Imaging of the Cosmic Assembly Near-infrared Deep Extragalactic Legacy Survey Fields (\uvc) survey is a \hst Cycle-26 Treasury Program (\hst-GO-15647, PI: Teplitz), awarded a sum of 164 orbits of primary Wide Field Camera 3 (WFC3) UVIS/F275W imaging with coordinated parallel Advanced Camera for Survey (ACS) F435W imaging, on four of the five premier extragalactic survey fields: GOODS-N, GOODS-S, EGS, and COSMOS.
The \uvc dataset is unique in further testing the LyC leakage bimodality mentioned above, since it is the uniform UV dataset surveying the largest area with UVIS, amounting to $\sim$426 arcmin$^2$, with extensive spectroscopic redshift measurements. The UV images across the entire fields, reaching a 5-$\sigma$ sensitivity of mag$_{\rm F275W}$=27 for compact sources, are taken at Hubble's angular resolution.
This is critical, since a key necessity in LyC searches is to confirm the origin of these faint signals with high spatial resolution imaging/spectroscopy, in order to exclude the possibility of foreground contamination from low-$z$ interlopers along the line of sight \citep[see \eg,][]{Steidel.2001,Iwata.2009,Vanzella.2010,Vanzella.2012,Nestor.2011kz8,Nestor.2013,Mostardi.2013}.

This paper is organized as follows. In Sect.~\ref{sect:uvc}, we first present a comprehensive overview of the key science drivers of the \uvc Treasury Program. In Sect.~\ref{sect:obs}, we present the observing strategy and data reduction details. We then explain the selection of our galaxy sample with spectroscopic redshifts in Sect.~\ref{sect:sample}. In Sect.~\ref{sect:stack}, we describe our image stacking techniques and show the results. The inference of the LyC \fesc is given in Sect.~\ref{sect:fesc}. Finally, we conclude in Sect.~\ref{sect:conclu}. Throughout this paper, the standard AB magnitude system is used \citep{Oke.1983}.

\section{Science overview of \uvc}\label{sect:uvc}

\uvc provides extensive UVIS imaging in four of the five premier \hst deep-wide survey fields: GOODS-N, GOODS-S, EGS, and COSMOS, targeted by the Cosmic Assembly Near-infrared Deep Extragalactic Legacy Survey \citep[\candels, ][]{Grogin.2011,Koekemoer.2011}. 
\uvc takes primary WFC3 F275W exposures at a uniform 3-orbit depth and coordinated parallel Advanced Camera for Survey (ACS) F435W exposures at slightly varying depth resulting from the roll angle constraints and the overlap from the increased FoV of the ACS camera. In total, the UV coverage secured by \uvc reaches $\sim$426 arcmin$^2$ \footnote{The number of pointings in each \uvc fields are 16 in GOODS-N, 8 in GOODS-S, 20 in EGS, and 16 in COSMOS.}, a factor of 2.7 larger than all previous data combined, including the WFC3 ERS UVIS imaging in GOODS-South \citep{Windhorst.2011}, the \candels F275W {Continuous Viewing Zone (CVZ)} imaging in GOODS-North \citep{Grogin.2011}, the UVUDF imaging in the Hubble Ultradeep Field within GOODS-South \citep{Teplitz.2013,Rafelski.2015}, and the HDUV covering portions of both GOODS fields \citep{Oesch.2018xn}.
This unique data set enables a wide range of scientific explorations as follows \citep{Wang.2020AAS}:
\begin{itemize}
    \item Using the high spatial resolution UV and Blue data (700 pc at $z\sim1$) to study the structural evolution of galaxies and create 2D maps of their star-formation history;
    \item Combining \uvc with the valuable Herschel legacy far-infrared data to trace the evolution of the star-formation, gas and dust content of moderate redshift ($z<1$) galaxies;
    \item Probing the role of environment in the evolution of low-mass ($\lesssim10^9\Msun$) star-forming galaxies at $z\sim1$;
    \item Investigating the decay of star-formation in massive early type galaxies and the role of minor mergers since $z\sim1.5$;
    \item Constraining the escape fraction of ionizing radiation (with $\lambda_{\rm rest}\le 912~$\AA) from galaxies at $z\gtrsim2.4$ to better understand how star-forming galaxies reionized the Universe at $z>6$.
\end{itemize}
In addition, this unique \uvc data set has additional treasury value: UV data break the degeneracy between the Balmer break and Lyman break spectral features \citep{Rafelski.2015} to improve the photo-z accuracy. HST UV data complement the existing and newly obtained ground-based U-band surveys in the \candels fields \citep{Ashcraft.2018,Redshaw.2022}.

\section{Observations and data reduction}\label{sect:obs}

\subsection{\uvc observations}\label{subsec:uvc_img}

\begin{figure*}[h!]
    \centering
    \includegraphics[width=\textwidth,trim=0cm 1cm 0cm 1cm,clip]{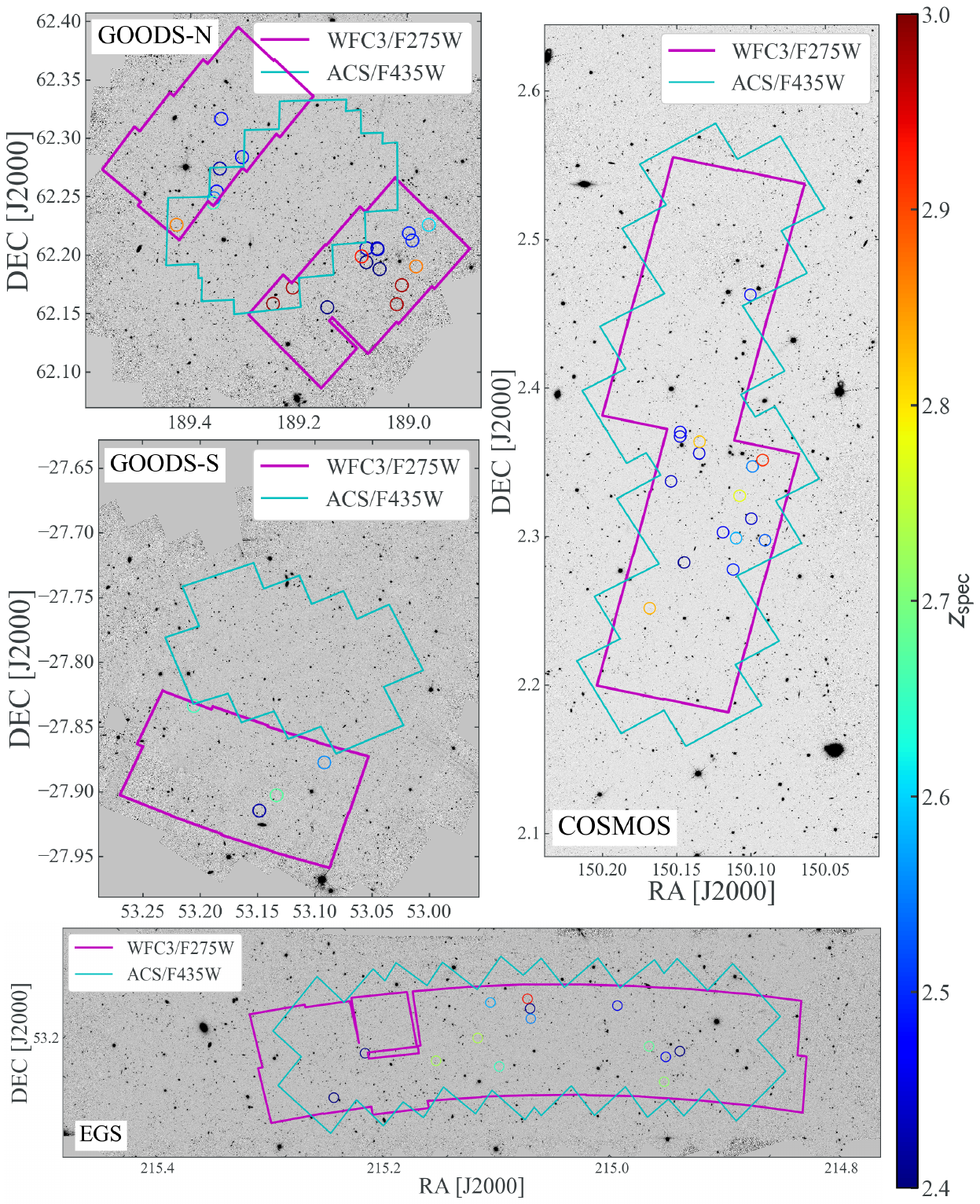}\\
    \vspace*{-.5em}
    \caption{\label{fig:uvcandels_fields}
    The \uvc survey footprints overlaid on top of the F814W mosaics of the four premier CANDELS fields: GOODS-N, GOODS-S, EGS, and COSMOS \citep{Grogin.2011,Koekemoer.2011}.
    Small chip gaps of the WFC3 and ACS detectors are not shown from the \uvc footprints.
    We also highlight the galaxies selected for the stacking analysis elaborated in Sect.~\ref{sect:stack}.
    They are color-coded in their spectroscopic redshifts visually inspected in 
    Sect.~\ref{sect:sample}. 
    The same color-coding is utilized in all four panels.
    }
\end{figure*}

The \uvc program (\hst-GO-15647, PI: Teplitz) obtained F275W and F435W (coordinated parallel) imaging of four of the CANDELS fields (see Fig. \ref{fig:uvcandels_fields}, providing new UV (F275W) and wide-area blue optical (F435W) coverage in the COSMOS and EGS fields, and doubling the UV area in GOODS. 

The survey was designed to reach a 5-$\sigma$ limiting magnitude of mag$_{\rm F275W}$=27 for compact galaxies (with $0\farcs2$ radius), which corresponds to an unobscured star-formation rate (SFR) limit of 0.2 \Msun/yr at $z\sim1$.
This goal required about 8100 seconds per pointing in F275W (3 orbits of two $\sim1350$ sec exposures each). A minimum of 6 exposures were required for good rejection of cosmic rays (CR), which was determined using the archival UVUDF \citet{Teplitz.2013} dataset.  As a result, \uvc did not choose the full-orbit exposures that are optimized for deep surveys with greater on-sky redundancy.  In GOODS-N, CVZ increased the efficiency of the observations. 

Following the Space Telescope Science Institute's best practices, exposures included post-flash to mitigate the effects of UVIS charge transfer efficiency (CTE) degradation \citep{Mackenty.2012}.  Post-flash protects against the loss of the faintest objects by filling ``traps'' on the charge-coupled devices (CCDs) before readout. The selected post-flash level brings the on-chip background up to 12 e$^-$ per pixel on average.

The F435W parallels have varying depth due to the overlap from the increased field of view (FOV) of the ACS camera. In COSMOS and EGS, {most of the area has 3} orbit depth with 5$\sigma$ sensitivity of mag$_{\rm F435W}$=28, while the overlap region have 6 orbit depth and a sensitivity of mag$_{\rm F435W}$=28.4. The GOODS fields already have B-band coverage of sufficient coverage (mag$_{\rm F435W}\sim$28), so the additional F435W data were placed in the central CANDELS-Deep region, where deep archival UV and near-infrared (NIR) data are available.

\uvc comprised 164 orbits of primary UVIS/F275W with ACS/F435W in parallel. Figure \ref{fig:uvcandels_fields} show the \uvc F275W and F435W footprints overlaid on the CANDELS optical imaging mosaics.  Small gaps between individual exposures in the F435W coverage are not shown in the figures.  The COSMOS and EGS were targeted for 16 and 20 pointings, respectively (approximately 2$\times$8 and 2$\times$10, adjusted for scheduling flexibility), enabling F435W coverage in parallel over most of the UV area.  In GOODS-S, \uvc obtained 8 pointings, to complete coverage of the field. In GOODS-N, \uvc observed two groups of 8 pointings ($2\times 4$), partially during CVZ and near-CVZ opportunities. The survey design employed a standard dithering pattern of \textsc{WFC3-UVIS-DITHER-LINE} to enable recovery of spatial resolution using \adriz to create mosaics (see Sect.~\ref{subsec:data_reduct} for more details).

In most cases, each pointing was observed in a single visit to enable robust CR rejection.  In a few cases, guide star limitations required splitting observations into 2-orbit and 1-orbit visits.   In 3 visits, initial observations failed and re-observations were obtained.  1 visit in the EGS fields failed after more than 90\% of the program had been completed, and so were not reobserved.

\subsection{\hst data reduction}\label{subsec:data_reduct}

The WFC3/UVIS and {ACS/WFC} images were calibrated using custom routines developed in \citet{Rafelski.2015,Prichard.2022}, and specifically for the \uvc data. In order to correct for radiation damage of the WFC3/UVIS detector over time, the F275W data were corrected with the updated charge transfer efficiency (CTE) correction algorithm \citep{Anderson.2021} which includes reduced noise amplification over previous efforts. These F275W images also include the official improved calibrations after 2021 that include concurrent dark subtraction to reduce the blotchy pattern otherwise observed in WFC3/UVIS images and updated flux calibration to match the latest \textsc{CALSPEC} models, which is especially important in the UV \citep{Calamida.2021}. We generate custom hot pixel masks using co-temporal darks and a variable threshold as a function of the distance to the readout to ensure a uniform number of hot pixels across the CCDs\footnote{\url{https://github.com/lprichard/HST_FLC_corrections}}. 

In addition, we flag readout cosmic rays (ROCRs) in the F275W data by identifying cosmic rays that fall on the detector after the readout of the amplifiers has begun, which are more apparent with the improved CTE correction code. The ROCRs appear as negative divots in the images due to over-correction by the CTE code, as the ROCRs land closer to the amplifiers than is recorded. The ROCRs are identified as 3 sigma negative outliers within 5 pixels of the readout direction of cosmic ray hits identified with AstroDrizzle. These negative pixels have the data quality flags set as bad pixels. Finally, we equalize the background levels on the four amplifiers as their bias level is likely affected by CTE degradation of the overscan regions. This results in clean images with constant background levels\footnote{\url{https://github.com/bsunnquist/uvis-skydarks}}. 

We also correct for scattered light in the WFC/ACS F435W images likely caused by earth limb light reflected off the telescope structure \citep{Biretta.2003,Dulude.2010}. Each exposure is checked for the existence of a gradient by comparing the left quarter to the right quarter of the image, and a difference threshold of 5e$^-$ is applied to a 3 sigma clipped median. If over the threshold, sources are masked in the image and the gradient is modeled with the \textsc{Photutils} Background2D module. Afterwards, the background level of the two chips is equalized to the mean level determined from a 3-$\sigma$ clipped mean after source masking\footnote{\url{https://github.com/bsunnquist/uvis-skydarks/blob/master/remove_gradients.ipynb}}.

\subsection{Image mosaicing}\label{subsec:mosaic}

The image registration and stacking pipeline was adapted from the one used in \citet{Alavi.2014}. Briefly, the calibrated, flat fielded WFC3/UVIS and ACS/WFC images were combined with the AstroDrizzle software package \citep{Gonzaga.2012}. To this end, {first}, we use the \textsc{Tweakreg} task in the \textsc{PYRAF/DrizzlePac} package to align the individual calibrated images within every single visit to ensure relative astrometric alignment. We then run the \adriz pipeline on each visit and align the drizzled output image to the CANDELS astrometric reference grid to correct for the small offset in pointing and rotation from different visits with different guide stars. {We performed the alignment on the 30 mas/pix images,} with a precision of 0.15 pixel, using unsaturated stars and compact sources.
These astrometric solutions are then transferred back to the header of the original calibrated flat fielded data using the \textsc{Tweakback} task in the \textsc{PYRAF/DrizzlePac} package. In the end, all of these aligned calibrated images are drizzled to the same pixel scale of {60 (30) milliarcsec} matched to the CANDELS reference images for the various fields\footnote{\url{https://archive.stsci.edu/hlsp/candels}}. Table~\ref{tab:drizzle_coef} provides the drizzle parameters that were used. The overall image footprints are shown in Fig.~\ref{fig:uvcandels_fields}.

\begin{deluxetable}{|l|l|}
    \label{tab:drizzle_coef}
    \tablecolumns{5}
    \tablewidth{0pt}
    \tablecaption{Values of some key \adriz parameters set in creating the new F275W and F435W mosaics from \uvc observations.}
    \tablehead{Parameter & Value}
\startdata
    \texttt{driz\_cr\_corr} & True \\
    \texttt{driz\_combine} & True \\
    \texttt{clean} & True \\
    \texttt{final\_wcs} & True \\
    \texttt{final\_scale} & 0.06 (0.03) \\
    \texttt{final\_pixfrac} & 0.8 \\
    \texttt{final\_kernel} & square \\
    \texttt{skymethod} & globalmin+match \\
    \texttt{skysub} & True \\
    \texttt{combine\_type} & imedian \\
\enddata
\end{deluxetable}

\adriz removes the background, rejects cosmic rays, and corrects input images for geometric distortion. In addition to the science images, \adriz generates an inverse variance map (IVM), which we use later to make weight images and to calculate photometry uncertainties. We apply the additional correction to the weight images to account for the correlated noise following \citet{Casertano.2000}. We make publicly available these image mosaics on MAST\footnote{\url{https://archive.stsci.edu/hlsp/uvcandels}}.

\section{Sample selections and spectroscopic redshifts}\label{sect:sample}

The focus of our LyC investigation is to measure or set strict limits on the escape fractions of LyC photons from galaxies at redshifts $z\gtrsim2.4$.  When the red end of the filter response is below the rest-frame Lyman limit, then detected signal can be interpreted as LyC emission \citep[see e.g.][]{Smith.2018,Smith.2020}. For this technique to succeed, it is important that the study relies on secure galaxy redshifts. In this section, we discuss the sample selection that enables our search for LyC escape.

\subsection{Redshift catalogs}\label{subsec:zcat}

To achieve precise, uncontaminated LyC flux measurements, we chose galaxies with high quality and accurate spectroscopic redshifts. To accomplish this, we vetted each spectrum in our spectroscopic sample by eye, then selected only the spectra with the highest score ranks by multiple spectroscopy experts on our team. 

Our initial compilation of archival spectral dataset comprises 
the Keck observations conducted by \citet{Barger.2008,Jones.2018log,Jones.2021loga}, 
the C3R2 survey \citep{Masters.2019},
the COSMOS-Magellan active galactic nucleus (AGN) survey \citep{Trump.2009zo},
the DEEP2 survey \citep{Newman.2013y9f},
the DEIMOS 10K survey \citep{Hasinger.2018},
the GOODS VLT/FORS2 survey \citep{Vanzella.2007},
the GOODS VLT/VIMOS survey \citep{Popesso.2009,Balestra.2010},
GMASS \citep{Kurk.2013},
the MOSDEF survey \citep{Kriek.2015},
the MUSE-Wide survey \citep{Herenz.2017}, 
\citet{Reddy.2006},
\citet{Szokoly.2004},
TKRS/TKRS2 \citep{Wirth.2004,Wirth.2015},
VANDELS \citep{Pentericci.2018e2s7}, 
VIPERS \citep{Garilli.2014}, 
VUDS \citep{LeFevre.2015}, 
VVDS \citep{LeFevre.2005}, and 
zCOSMOS \citep{Lilly.2007}. 

We also supplement this extensive redshift catalog with additional sources identified in a reanalysis of archival \hst near-infrared grism spectroscopy in the \candels fields, acquired by primarily the 3D-HST survey \citep{Brammer.2012,Momcheva.2016lbw}.
This reanalysis is part of a novel initiative dubbed the Complete Hubble Archive for Galaxy Evolution \citep[CHArGE, see \eg,][]{Kokorev.2022}. The state-of-the-art \grzl software\footnote{\url{https://github.com/gbrammer/grizli}} \citep{Brammer.2021} is utilized to conduct uniform reprocessing of all archival \hst imaging and slitless spectroscopy in the areas of \candels fields contained within our \uvc footprints. 

Briefly, \grzl reduces the paired pre-imaging and grism exposures in five steps.
A) Pre-processing the raw WFC3 imaging and grism exposures. This step includes flat fielding, relative/absolute astrometric alignment, {variable/master sky background removal}, satellite trail masking, etc.
B) Iterative forward-modeling the full field-of-view (FoV) grism exposures at visit level.
C) Obtaining best-fit grism redshift of sources through spectral template synthesis.
D) Refining the full FoV grism models.
E) Extracting the science-enabling products (\eg 1D/2D grism spectra, emission line maps, etc.) for sources of interest.
During step C), \grzl calculates a number of goodness-of-fit statistics for the redshift fitting procedure. As in \citet{Wang.2022a,Wang.2022b}, we compiled a list of galaxies with secure grism redshifts at $z_{\rm grism}\gtrsim2.4$, when all of the following goodness-of-fit criteria are satisfied:
$\chi^2_{\rm reduced}<1.5 \wedge \left(\Delta z\right)_{\rm posterior}/(1+z_{\rm peak})<0.005 \wedge \mathrm{BIC}>30$.
Here $\chi^2_{\rm reduced}$ is the reduced $\chi^2$, $\left(\Delta z\right)_{\rm posterior}$ represents the 1-$\sigma$ width of redshift posteriors, and BIC stands for the Bayesian information criterion estimated from the template fitting procedure.
This provides us additional high-quality grism redshifts that we combine with the previous redshift compilations for the subsequent visual inspection, to select the highest fidelity sample for our stacking analysis.
In addition, this coherent reanalysis of {all the available archival} grism exposures covering the entire \uvc fields provides comprehensive measurements of the nebular emission line properties of our sample galaxies, which help divide the full sample into subgroups (see Sect.~\ref{subsec:inspect}).

\subsection{Visual inspection and sample selection}\label{subsec:inspect}

For our visual inspection, we assigned each spectrum to be visually vetted by at least four reviewers randomly chosen from a group of 13 spectroscopy experts. Each reviewer gave a score to the claimed redshift from the survey catalog, with the following rubric, \ie, 0: reliable, 1: possibly correct, or 2: incorrect. This ranking takes into account the alignment of the observed spectral features to their expected position given the reported redshift in the spectrum, the amount of visible noise, the shape of each line or other spectral feature, the presence of unknown/unexpected spectral lines, the presence of neighbors in the corresponding multiband \hst images, and the drop-out band from the \hst images. After ranking was completed, we separated each spectrum into five redshift quality quintiles, based on their median and average scores, where the first quintile has median<1 and average<0.5, the second quintile has median<1 and average<1, the third quintile has median<2 and average<1.5, the fourth quintile has average<2 and median<2, and the last quintile has median=2 and average=2.
As a consequence, our first redshift quality quintile has 39 galaxies, second has 59, third has 202, fourth has 19, and fifth has 271. Example figures of galaxies in each quintile are shown in the Appendix \S\ref{sec:appendixInspect}.

In our analysis, we will rely on the most secure redshifts, referring to those in the first, second and third redshift quality quintiles identified by our reviewers. 
We base our decision to include the third quintile on experience with the overall high quality of the spectra in \candels.
In this paper, we focus on the search of the LyC signals from individual leakers and/or stacking analysis using the \uvc WFC3/F275W imaging data. So we trim the high-quality redshift catalog by filtering out all spectra taken outside of the \uvc footprints (we defer the analysis outside the \uvc footprints but within other archival UV imaging to a later work).
We identify and exclude potential AGN candidates from the deep Chandra X-ray observations in these \candels fields \citep[][, and D. Kocevski priv. comm.]{Hsu.2014,Nandra.2015,Xue.2016}, as we present the stacking analysis of the escaping LyC signals from AGNs in a companion work \citep{smithLymanContinuumEmission2024a}.
We also exclude sources that lie in close proximity to detector chip gaps of the \uvc F275W imaging, for complex noise properties due to insufficient dithering.

In the end, we compiled a list of \Ntot galaxies with high-quality spectroscopic redshifts in the range of $z\in[2.4,3.0]$, based on our dedicated visual inspections of these publicly available spectra.
Among these \Ntot galaxies, \Ndet show significant detection of F275W fluxes with a signal-to-noise ratio (SNR) $\geq$3, whose detailed multi-wavelength photometric measurements are presented in Appendix \S\ref{appendix:indvd_lyc}. We do not opt to include them in the stacking to avoid stacks dominated and biased by a few sources which can likely be extreme outliers.
All the rest \Nstack galaxies in our high-quality spectroscopic redshift sample show non-detections in F275W, and are analyzed in our stacking analysis elaborated in Sect.~\ref{subsect:stack}.

To take into account the rapidly evolving IGM opacity with respect to redshift, we divide the entire \Nstack galaxies into two redshift samples: the \lowz bin ($z\in[2.4,2.5]$) and the \medz bin ($z\in[2.5,3.0]$), consisting of \Flowz and \Fmedz galaxies, respectively (the ``full'' samples).
Galaxies in the \lowz bin have their Lyman limit lying closer to the red edge of F275W than sources in the \medz bin.
Within both redshift bins, we also define four sub-groups on which the stacking analysis is performed separately, for the purpose of testing the possible correlation between LyC leakage and galaxy global properties.
First and foremost, we classify all the galaxies in the first and second redshift quality quintiles with median score <1 and average score <1 (\ie at least 3 out of 5 inspectors deem this redshift measurement secure) as the \goldz subsamples.
Then, we refer to the galaxies showing significant detection of \OII$\lambda$3727,3730 (hereafter referred to as \OII) emission in their WFC3/G141 grism spectra as the strong line emitter subsample, denoted by \snOII.
Moreover, galaxies with intrinsically faint absolute UV magnitude ($M_{\rm UV}$>-20.5) are selected to form the \hiMuv subsamples.
Last but not the least, galaxies with poor dust content (E(B-V)$<$0.2) are considered to be the \lowEbv subsamples.
In total, we have 10 stacking samples in two mutually exclusive redshift ranges. The break-down of the number counts for the full and sub samples analyzed in our stacking analysis are presented in Table~\ref{tab:source_cnt}. 

{
\tabcolsep=2pt
\begin{deluxetable*}{cccccccccccccccccc}
    \label{tab:source_cnt}
    \tablecolumns{6}
    \tablewidth{0pt}
    \tablecaption{Number of sources with high-quality spectroscopic redshifts at $z\in[2.4,3.0]$, compiled from our detailed efforts visually inspecting publicly available archival spectroscopic observations (see Sect.~\ref{subsec:inspect}). The entire spectroscopic sample consists of 56 galaxies in total, all of which show non-detections (SNR$<$3) in the \uvc F275W imaging data.
    These 56 galaxies are further separated into two redshift bins to facilitate direct comparison of the LyC escape fraction, given the redshift evolution of the IGM transmission and rest-frame wavelength coverage by F275W. For each redshift bin, we further extract sub-samples based on specific physical properties of the galaxies. The stacking analysis is applied to the full and sub samples (see Sect.~\ref{subsect:stack}).}
\tablehead{
     Redshift\tablenotemark{a} & Full & \multicolumn{4}{c}{Sub-sample} \\
     Bin & Sample    & \multicolumn{4}{c}{\hrulefill}  \\
     &  & \goldz\tablenotemark{b} & \snOII\tablenotemark{c} & \hiMuv\tablenotemark{d}  & \lowEbv\tablenotemark{e}  \\
}
\startdata
 \lowz    &  28 &  12 &  11 &  11 &  13 \\
 \medz  &  28 &  7   &  15 &  14 &  19 \\
\enddata
    \tablenotetext{a}{The \lowz and \medz redshift bins bracket galaxies in the redshift range of [2.4,2.5] and [2.5,3.0], with a median redshift of $z_{\rm median}=2.44$ and $z_{\rm median}=2.72$, respectively, for the full samples.}
    \tablenotetext{b}{The \goldz sub-sample includes galaxies with the most secure spectroscopic redshifts, compiled from our visual inspection efforts.}
    \tablenotetext{c}{The \snOII sub-sample showing significant \OII emission with SNR$_{\OII}$ $\gtrsim$ 3, measured from their archival G141 spectra. These galaxies are considered to be prominent line emitters, highly likely those leaking copious amount of LyC flux.}
    \tablenotetext{d}{The \hiMuv sub-sample refers to galaxies with intrinsically faint absolute UV magnitude of $M_{\rm UV}>-20.5$.}
    \tablenotetext{e}{The \lowEbv sub-sample corresponds to dust-poor galaxies with ${\rm E(B-V)}<0.2$.}
\end{deluxetable*}
}

\section{Image stacking and photometry}\label{sect:stack}

In this section, we first introduce the procedures to perform spectral energy distribution (SED) fitting to the stacking sample in Sect.~\ref{subsect:sed}.
Then we describe in detail our methodology of image stacking in Sect.~\ref{subsect:stack}.
Finally we present our aperture photometry method of the stacked images in Sect.~\ref{subsect:photom}.

\subsection{SED fitting to the stacking sample}\label{subsect:sed}

We conduct UV-optimized aperture photometry on the new WFC3/F275W and ACS/F435W \uvc data, following the methodology developed as part of the Hubble Ultra-Deep Field UV analysis \citep{Teplitz.2013,Rafelski.2015}.
This method regards the object optical isophotes as more appropriate apertures for counting UV photons than the isophotes defined in near-infrared wavelengths.
The PSF and aperture corrections are then performed to make sure the UV and B-band fluxes are consistent with the previous measurements at other wavelengths.
The detailed description of our UV-optimized aperture photometry method is described in \citet{sunUltravioletLuminosityFunction2024}.

We perform detailed Bayesian inference of the stellar population properties for all the \Nstack galaxies in the stacking sample, using the CIGALE software \citep{Boquien.2019} to fit their multi-wavelength photometric measurements, the majority of which are taken from the public CANDELS photometric catalogs, as published in \citet{Barro.2019} and \citep{Stefanon.2017} for GOODS-N and EGS, respectively.
We model the galaxy star-formation history (SFH) using the delayed $\tau$ model via the \textsc{sfhdelayed} module, since the sudden onset of star formation and burst episodes in a, e.g., double-exponential parameterization (\ie the \textsc{sfh2exp} module) may be too abrupt when the variation of the SFH may be smoother \citep{Boquien.2019,Carnall.2019}.
We also verify that the different choices of SFH models have a marginal effect on the estimated \fesc values, on average by $\sim$4\%.
We rely on the BC03 stellar population synthesis models \citep[the \textsc{bc03} module,][]{Bruzual.2003}, the infrared dust models of \citet{Dale.2014}, and the \citet{Calzetti.2000} dust extinction law, during the spectral energy distribution (SED) fitting with CIGALE.
In particular, we utilize the \textsc{nebular} module and keep the parameter \textsc{nebular.f\_esc} freely varying to allow for emission lines in the SED.

\subsection{Stacking methodology}\label{subsect:stack}

\begin{figure*}[h!]
    \centering
    \includegraphics[width=.9\textwidth,trim=0cm 0cm 0cm 0cm,clip]{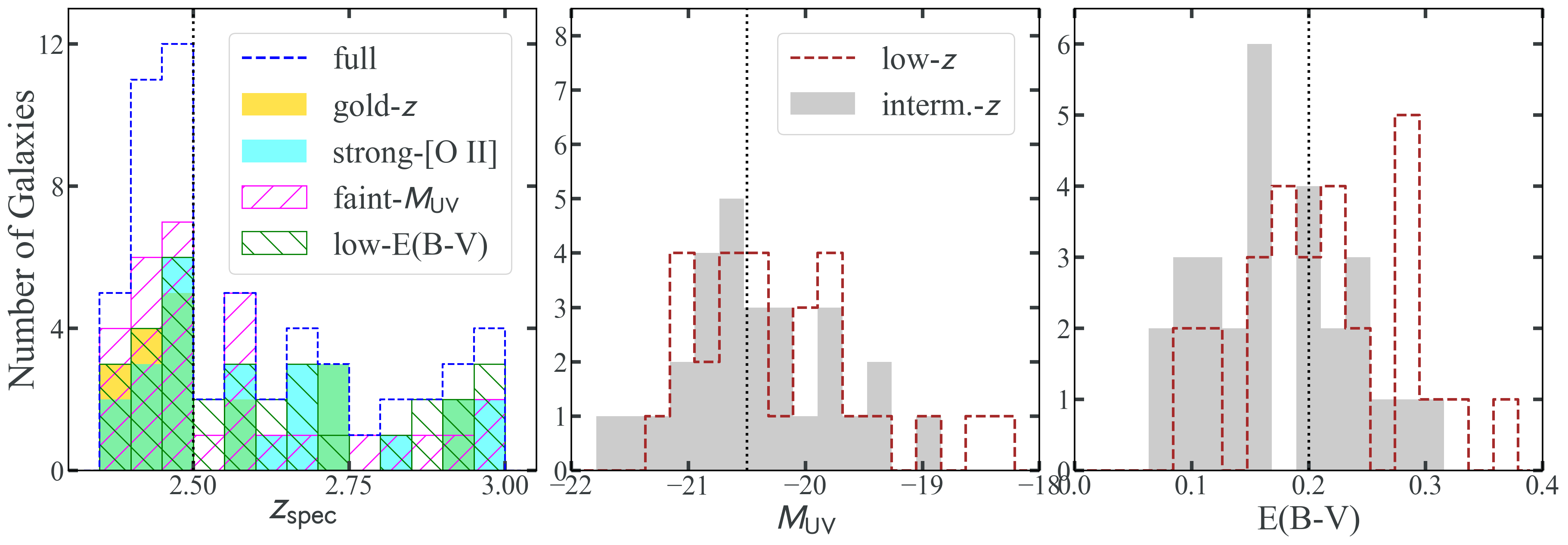}\\
    \vspace*{-.5em}
    \caption{\label{fig:histograms}
    Distribution of spectroscopic redshifts ($z_{\rm spec}$), absolute UV magnitude ($M_{\rm UV}$) and dust attenuation (E(B-V)) of the selected stacking samples.
    To facilitate direct comparison of the LyC escape fraction, all stacked galaxies are separated into two redshift bins of [2.4,2.5] (\lowz) and [2.5,3.0] (\medz), marked by the black vertical dotted line in the left panel.
    The black vertical dotted lines in the middle and right panels correspond to the demarcation values of $M_{\rm UV}>-20.5$ and E(B-V)$<$0.2, used to compile the \hiMuv and \lowEbv subsamples, respectively.
    The numbers counts of the full and sub-samples within the two redshift bins are broken down in Table~\ref{tab:source_cnt}.
    }
\end{figure*}

Our entire stacking sample consists of \Nstack galaxies with secure spectroscopic redshifts, all showing non-detections (\ie SNR<3) in our \uvc F275W imaging.
Following the SED fitting procedures outlined in Sect.~\ref{subsect:sed}, we derive the estimates of their absolute UV magnitudes ($M_{\rm UV}$) and dust attenuation values (E(B-V)) shown as the histograms in Fig.~\ref{fig:histograms}. As given in Table~\ref{tab:source_cnt}, we separate these \Nstack galaxies into two redshift bins, and obtain in total 10 individual stacking samples, to constrain the LyC escape fraction on a population level, taking into account the rapidly evolving LyC opacity in the IGM \citep{Steidel.2018,bassettIGMTransmissionBias2021}.
We adopt the following image stacking procedures, similar to those utilized in \citet{Smith.2018,Alavi.2020,Smith.2020,smithLymanContinuumEmission2024a}.
Consistent with the previous studies of LyC leakage that employ stacking procedures, we assume the galaxies selected in our sample to be self-similar, thus allowing us to stack them to make a more accurate measurement of the faint LyC flux.

\begin{figure*}[t!]
    \centering
    \includegraphics[width=.9\textwidth,trim=0cm 11cm 0cm 4.3cm,clip]{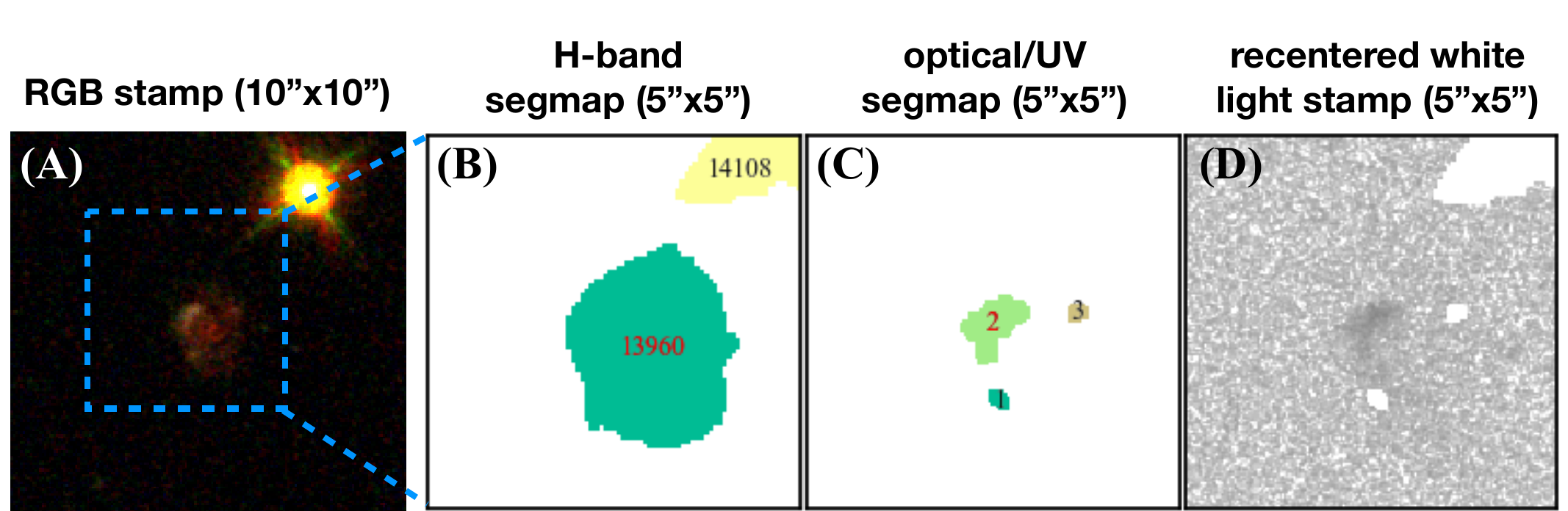}\\
    \vspace{-.5em}
    \caption{\label{fig:recenter}
    An illustration of our re-centroiding and contaminant-masking methodology via performing stamp photometry before the stacking analysis.
    {\bf (A)} galaxy color-composite stamp with size $10\arcsec\times10\arcsec$ made from the \candels near-infrared and optical imaging, centered on the F160W light centroid. The blue dashed square ($5\arcsec\times5\arcsec$) marks the spatial extent of panels {\bf (B)-(D)}, which are all the same.
    {\bf (B)} \candels F160W segmentation map defined using the hot+cold source detection scheme \citep[see \eg,][]{Guo.20139ts}. The ID of our source of interest is marked in red, while the ID of a neighbouring source (a star spike) in black.
    {\bf (C)} optical/UV segmentation map using the white light image as detection image, while masking all neighbouring sources. The region \#2 has the brightest optical flux, so the light centroid of region \#2 is used to re-center panels {\bf (B)-(D)}.
    {\bf (D)} the white light image stamp with neighbouring sources and segmented regions within our source of interest (\ie regions \#1 and \#3) masked.
    Here we clearly see that the light centroids of optical/UV and near-infrared bands do not overlap. These possible offsets can be effectively accounted for using our re-centroiding and contaminant-masking methodology.
    }
\end{figure*}

\begin{enumerate}
    \item We first make large cutout image stamps in multiple filters (F275W and F435W from our work described in Sect.~\ref{subsec:mosaic}, F606W and F814W from \candels) with size $10\arcsec\times10\arcsec$ on 0\farcs03 plate scale, centered on each galaxy with secure spectroscopic redshift that passes our visual vetting described in \ref{subsec:inspect}.
    {We choose the coadded mosaics on 0\farcs03 plate scale to take full advantage of the high resolution of \hst imaging.}
    Yet note that at this stage the object coordinates (RA and Dec) are taken from the CANDELS/UVCANDELS multi-wavelength photometric catalogs that use \H-band mosaics as the detection image \citep{Guo.20139ts,Nayyeri.2017bnk,Stefanon.20172qt,Barro.2019}.  The light centroids measured in F160W and optical/UV filters often do not align, as they probe the rest-frame optical and UV wavelength ranges of our sources of interest.

    \item To properly re-centroid the optical/UV image cutouts before stacking and {at the same time} mask possible neighboring contaminants, we utilize the \textsc{Photutils} software to perform photometry on these cutout images.
    We produce the white light image from the image cutouts in all available filters weighted by filter mean flux density (\ie $f_{\nu}$, the \textsc{PHOTFNU} keyword), which is taken as the detection image.
    We mask regions of pixels in the image cutouts according to existing object isophotes already defined in the \H-band segmentation maps from the \candels photometry.
    
    \item After obtaining the segmentation map from the detection white light image cutout, we measure the isophotal fluxes associated with each of the segmentation regions in the individual filter image cutouts using \textsc{Photutils}, and select the region that has the brightest {F606W} flux, under the assumption that the ionizing radiation emerges from the areas dominated by the emission from young massive stars. This is critical in pinpointing the centroid of our galaxies' rest-frame UV light, where the LyC signal most likely originates.
    
    \item We then make small cutout image stamps with $5\arcsec\times5\arcsec$, appropriately centered on the object's peak {\V-band flux (corresponding to object's UV-light-weighted centroid)}, with all contaminants masked. Fig.~\ref{fig:recenter} demonstrates the entire procedures for our re-centroiding and masking strategy. 
    We also estimate the local sky background and remove that from the SCI extension before stacking. We exclude all surrounding objects outside a certain circular aperture of 0.5 arcsec.
    
    \item To combine the surface brightness signals from small cutout image stamps of individual objects, we adopt the UV luminosity ($L_{\rm UV}$) normalization, described by
    \begin{align}\label{eq:Luv_wht}
        {\bar f} = \frac{\Sigma(f_i/L^{i}_{\rm UV})}{N_{\rm obj}}\cdot L^{\rm median}_{\rm UV}, \qquad \sigma^2 = \frac{\Sigma \left(\sigma_i/L^{i}_{\rm UV}\right)^2}{N^2_{\rm obj}}\cdot (L^{\rm median}_{\rm UV})^2,
    \end{align}
    where $f_i$ and $\sigma_i$ stand for the flux and uncertainty of the $i^{\rm th}$ source among the entire number of $N_{\rm obj}$ that contribute to the stacks. $L^{i}_{\rm UV}$ represents the galaxy rest-frame UV luminosity converted from $M_{\rm UV}$ shown in Fig.~\ref{fig:histograms}. $\bar f$ and $\sigma$ thus denote the stacked flux and uncertainty.
  
    Eq.~\ref{eq:Luv_wht} normalizes galaxy multi-band photometry before stacking as the LyC escape fraction is a relative quantity \citep{Marchi.2017,Steidel.2018}.
    In Sect.~\ref{sect:fesc}, we show results obtained from this stacking method.
    
    \item Finally, we apply the $L_{\rm UV}$ normalization to all galaxy samples listed in Table~\ref{tab:source_cnt}. The large spatial coverage of the \uvc survey presents us an ideal opportunity of testing the strengths of escaping LyC signals as a function of galaxy global properties with sufficient sample statistics. The stacked images in multiple filters for the \medz full sample consisting of \Fmedz sources are presented in Fig.~\ref{fig:stack_cnt}.

\end{enumerate}

\subsection{Photometry on image stacks}\label{subsect:photom}

Before image stacking detailed in Sect.~\ref{subsect:stack}, we have already subtracted a constant from each cutout image stamp in each filter (F275W, F435W, F606W, and F814W) of each galaxy selected for stacking, to ensure that the mode of the local sky background of each galaxy is achromatically as close to zero as possible.
Just in case there still remains some residual sky background signals that survive our stacking procedures, we perform another round of background estimation and removal before the photometry of stacked images.
We apply a circular mask with a radius of 45 pixels (\ie 1\farcs35) to the central regions of the stacked images in each filter, and bin the unmasked background pixels according to the Freedman-Diaconis rule, following \citet{Smith.2018,Smith.2020}.
Then we derive the mode value of the resultant count-rate histogram of these surrounding pixels outside the central masked regions, as the remaining sky background in the stacked images.
After subtracting off this remaining background, we make sure that the empty regions of our stacked images only contain random noise.

The sample median redshifts are given in Table~\ref{tab:fesc_stack}. At these redshifts, the broad-band F606W filter probes the bulk of the rest-frame 1500 \AA\ wavelength regime of source spectrum, conventionally regarded as the UV continuum (UVC).
The escaping LyC signals are believed to originate from young star clusters comprising hot, massive O-type stars, which also dominate the blue-UV wavelength range of the host galaxy's SED \citep[see \eg,][]{Vanzella.2022}.
We therefore take the stacked F606W image as the detection image to define the segmentation map for flux measurements as shown in Fig.~\ref{fig:stack_cnt}.
Following the practice presented in Sect.~\ref{subsect:stack}, we utilize the \textsc{Photutils} software to measure the isophotal fluxes in the stacked images in all four filters: F275W, F435W, F606W, and F814W.
We adopt the following parameters for \textsc{Photutils}:
\texttt{detect\_thresh}=1, \texttt{analysis\_thresh}=1, \texttt{detect\_minarea}=8,
\texttt{deblend\_nthresh}=32,\\ \texttt{deblend\_mincont}=0.001.
\textsc{Photutils} reports the sum of fluxes within the F606W isophot for all four filters.
Finally, using the up-to-date zero points of these four filters, we convert the measured fluxes into isophotal magnitudes given in Table~\ref{tab:fesc_stack}.
After averaging the best-fit CIGALE SEDs as discussed above, we find an insignificant red-leak of flux into the F275W filter redward of the 912\AA\ Lyman break of $\sim$\,0.8\%.

{
\tabletypesize{\scriptsize}
\tabcolsep=2pt
\begin{deluxetable*}{lccccccccccccccccc}
    \label{tab:fesc_stack}
    \tablecolumns{15}
    \tablewidth{0pt}
    \tablecaption{Measured photometry and physical properties of the selected galaxy samples from our stacking analyses using $L_{\rm UV}$ normalization.}
\tablehead{
    Redshift &  Sample & $N_{\rm gal}$ & $z_{\rm median}$\tablenotemark{a} &  \multicolumn{5}{c}{Photometric measurements from the stacked images\tablenotemark{b}}  &  \multicolumn{6}{c}{Measured physical properties}  \\
    Bin &  &  &  & \multicolumn{5}{c}{\hrulefill}  & \multicolumn{6}{c}{\hrulefill}  \\
     &  &  &  &  
        mag$_{\rm F275W}$  & 
        mag$_{\rm F435W}$  & 
        mag$_{\rm F606W}$  & 
        mag$_{\rm F814W}$  & 
        $\left( \frac{F_{\rm LyC}}{F_{\rm UVC}} \right)_{\rm obs}$\tablenotemark{c}  &
        $M_{\rm UV}$ &
        E(B-V) &
        $\left( \frac{F_{\rm UVC}}{F_{\rm LyC}} \right)_{\rm int}$\tablenotemark{d}  &
        $\langle\bar t_{\rm IGM}\rangle$\tablenotemark{e} [\%] &
        $\fesc^{\rm rel}$\tablenotemark{f} [\%] &
        $\fesc^{\rm abs}$\tablenotemark{f} [\%]
}
\startdata
\lowz  &  full & 28 & 2.44 & $>$28.70 & 25.44 & 24.74 & 24.46 & $<$0.026 & -20.41 & 0.21 & 5.06 & 36.5 [30.0, 42.9] & $<$36.8 & $<$5.0 \\ 
\lowz  &  goldz & 12 & 2.45 & $>$28.99 & 25.13 & 24.59 & 24.40 & $<$0.017 & -20.57 & 0.19 & 4.98 & 37.0 [27.1, 46.8] & $<$23.7 & $<$3.7 \\ 
\lowz  &  snOII & 11 & 2.46 & $>$28.66 & 25.23 & 24.59 & 24.45 & $<$0.024 & -20.54 & 0.21 & 4.58 & 35.7 [25.6, 45.9] & $<$32.6 & $<$4.4 \\ 
\lowz  &  hiMuv & 17 & 2.44 & $>$28.71 & 26.11 & 25.27 & 24.90 & $<$0.042 & -19.88 & 0.24 & 6.11 & 36.9 [28.5, 45.1] & $<$71.1 & $<$7.1 \\ 
\lowz  &  lowEbv & 13 & 2.44 & $>$28.50 & 25.02 & 24.49 & 24.42 & $<$0.025 & -20.68 & 0.16 & 4.58 & 36.6 [27.1, 46.1] & $<$32.4 & $<$7.0 \\ 
\hline\noalign{\smallskip}
\medz  &  full & 28 & 2.72 & $>$28.61 & 25.84 & 25.04 & 24.80 & $<$0.037 & -20.52 & 0.15 & 6.07 & 20.0 [14.5, 25.6] & $<$113.1 & $<$26.1 \\ 
\medz  &  goldz & 7 & 2.73 & $>$28.01 & 25.42 & 24.68 & 24.51 & $<$0.046 & -20.84 & 0.23 & 6.82 & 20.2 [8.9, 31.4] & $<$201.1 & $<$22.5 \\ 
\medz  &  snOII & 15 & 2.71 & $>$28.21 & 25.58 & 24.80 & 24.66 & $<$0.043 & -20.73 & 0.16 & 6.00 & 20.1 [12.5, 27.6] & $<$133.6 & $<$29.0 \\ 
\medz  &  hiMuv & 14 & 2.64 & $>$28.73 & 26.56 & 25.67 & 25.34 & $<$0.059 & -19.92 & 0.18 & 6.46 & 22.0 [13.7, 30.2] & $<$193.6 & $<$35.4 \\ 
\medz  &  lowEbv & 19 & 2.68 & $>$28.47 & 25.60 & 24.97 & 24.82 & $<$0.040 & -20.60 & 0.14 & 5.86 & 20.3 [13.4, 27.0] & $<$117.5 & $<$32.5 \\ 
\enddata
    \tablenotetext{a}{Sample median redshift.}
    \tablenotetext{b}{Measurements and 1-$\sigma$ limits for the isophotal magnitudes and flux ratios derived from the stacked image stamps. The magnitudes are in unit of ABmag.}
    \tablenotetext{c}{The 1-$\sigma$ upper limits of the observed ratio of the escaping LyC flux versus the UVC flux, measured from the empty aperture analysis performed on the stacked F275W images (see Sect.~\ref{sect:fesc}). For each stack, we take the 84.13 percentile of the distribution resulting from the empty aperture analysis as the 1-$\sigma$ upper limit reported here, following \citet{gehrelsConfidenceLimitsSmall1986}.} 
    \tablenotetext{d}{The intrinsic ratio of the UVC and LyC flux. We compute this ratio for each galaxy within the stacking samples based on their best-fit intrinsic SED and take the median of the distribution as the intrinsic ratio for stacks, listed here.}
    \tablenotetext{e}{The mean IGM transmission and 1-$\sigma$ confidence interval drawn from the average IGM transmission distribution of all galaxies in the respective stacks. We simulate 10,000 sightlines for each galaxy using the \tao Monte Carlo line-of-sight IGM code \citep{bassettIGMTransmissionBias2021}, to account for the high stochasticity of IGM transmission (see Fig.~\ref{fig:IGMtrans}).}
    \tablenotetext{f}{The 1-$\sigma$ upper limits of the relative and absolute escape fractions, calculated using 3,000 bootstrap random draws from the mean IGM transmission distribution of each stack and 3,000 observed LyC-to-UVC flux ratios measured from the empty aperture analysis of each stack.
    We adopt $\fesc^{\rm abs} = \fesc^{\rm rel} \, 10^{-0.4\,A_{\rm UV}}$, where $A_{\rm UV}=10.33\,{\rm E(B-V)}$ is the \citet{Calzetti.2000} dust reddening law. For each stack, we take the 84.13 percentiles of the resulting histograms of $\fesc^{\rm rel}$ and $\fesc^{\rm abs}$ as their 1-$\sigma$ upper limits reported here, following \citet{gehrelsConfidenceLimitsSmall1986}.
    }    
\end{deluxetable*}
}

\section{Inference of the LyC escape fraction}\label{sect:fesc}

In this section, we present the main results from our detailed stacking analysis. Following the standard nomenclature \citep[\eg,][]{Steidel.2001,Siana.2010,Siana.2015,Steidel.2018}, we define the \emph{relative} escape fraction of LyC as 
\begin{align}\label{eq:fesc_rel}
    \fesc^{\rm rel} = \frac{ \left( F_{\rm UVC}/F_{\rm LyC} \right)_{\rm int} }{ \left( F_{\rm UVC}/F_{\rm LyC} \right)_{\rm obs} } \cdot \exp(\tauIGM),
\end{align}
where $\left( F_{\rm UVC}/F_{\rm LyC} \right)_{\rm int}$ and $\left( F_{\rm UVC}/F_{\rm LyC} \right)_{\rm obs}$ represent the intrinsic and the observed ratios of the UVC and LyC fluxes, respectively, and $\tauIGM$ is the IGM opacity.
The transmission of the ionizing flux through the intervening IGM (\ie $t_{\rm IGM}=\exp(-\tauIGM)$) is a highly stochastic process, ascribed to its bimodal probability distribution function (PDF). \citet{bassettIGMTransmissionBias2021} showed that the PDF of $t_{\rm IGM}$ can be depicted by a sudden rise towards $t_{\rm IGM}\sim0$ and a much less prominent second peak at higher values. Due to the strong degeneracy between \fesc and $t_{\rm IGM}$, the inferred values of \fesc are often overestimated from the observations of galaxies leaking strong LyC flux under the assumption of a mean $t_{\rm IGM}$, a.k.a. the IGM transmission bias. Fortunately, \citet{bassettIGMTransmissionBias2021} verified that this bias is less fatal for large galaxy ensembles including LyC non-detections. 

\begin{figure*}[h!]
    \centering
    \includegraphics[width=\textwidth,trim=0cm 0cm 0cm 0cm,clip]{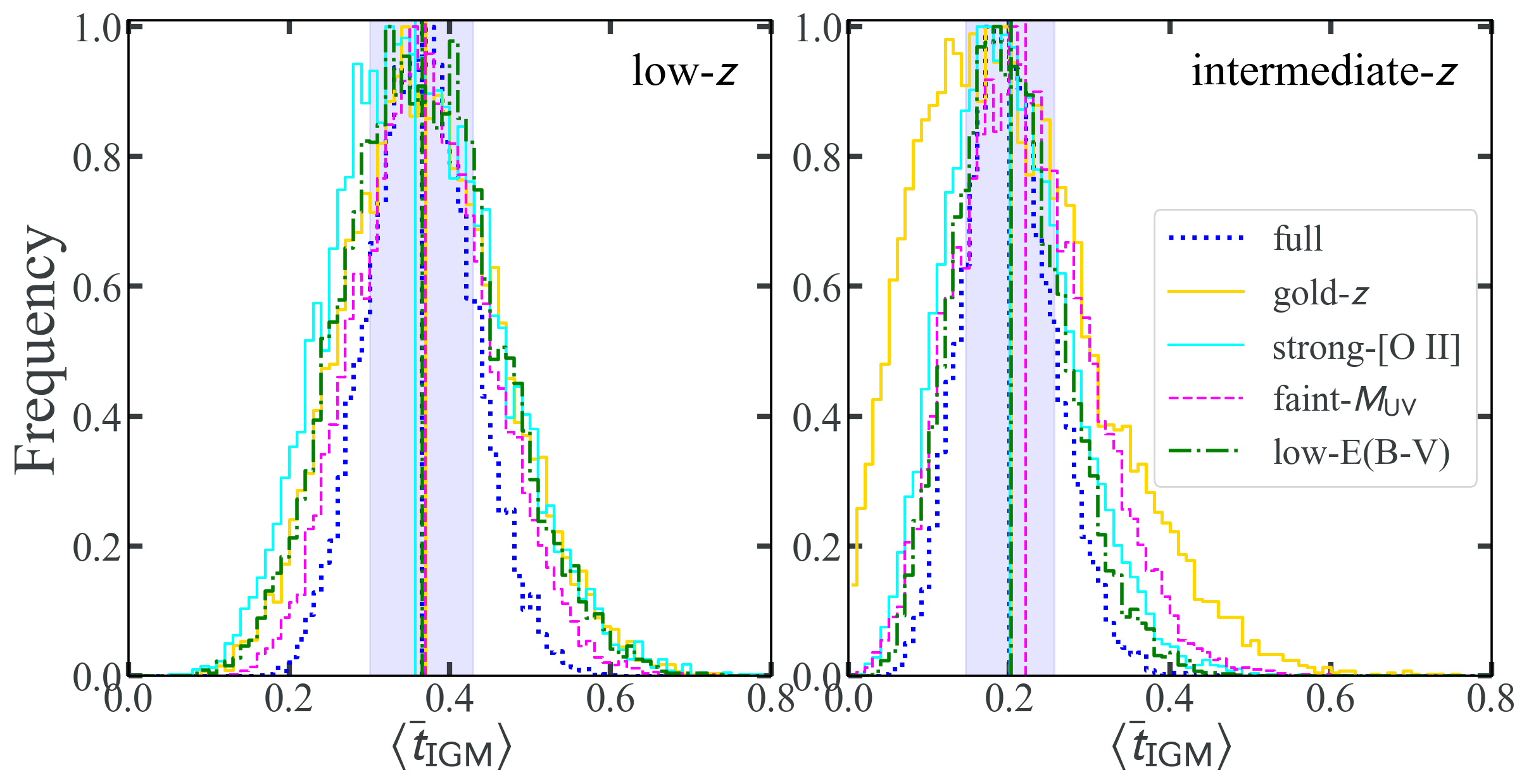}
    \vspace*{-1em}
    \caption{\label{fig:IGMtrans}
    Mean IGM transmission distributions for each galaxy sample in the two redshift bins: \lowz and \medz shown in the left and right panels, respectively.
    In both panels, we show the mean IGM transmission histogram averaged from the full sample in blue dotted lines, with the histograms for the four subsamples represented in other lines.
    The vertical straight lines correspond to the median values for each histogram, and the blue shaded region marks the 1-$\sigma$ range for the full sample. 
    We see that averaging the IGM transmission values randomly drawn from bimodal distributions for a relatively large sample of galaxies at similar redshifts lead to a single-peaked mean IGM transmission distribution, useful for breaking the degeneracy between $t_{\rm IGM}$ and the \fesc inference for the galaxy sample \citep{naiduLowLymanContinuum2018,bassettIGMTransmissionBias2021}.
    }
\end{figure*}

To determine the appropriate average IGM transmission value for each stack, we use the \tao software\footnote{\url{https://github.com/robbassett/TAOIST_MC}} \citep{bassettIGMTransmissionBias2021}
to generate 10,000 realizations of the line-of-sight IGM transmission models as a function of observed wavelength per galaxy at its known redshift. We then calculate the filter throughput weighted average of these transmission models for each line-of-sight using
\begin{align}\label{eq:t_IGM}
    \bar t_{\rm IGM} = \frac{\int \exp(-\tauIGM)\, \frac{T_{\rm F275W}}{\lambda} \, \ud\lambda}{\int \frac{T_{\rm F275W}}{\lambda} \, \ud\lambda}
\end{align}
with $T_{\rm F275W}$ representing the F275W filter throughput.
This ensures that only the observed wavelengths within the F275W filter are considered in the IGM transmission calculations, and any wavelengths outside of the filter would not be used to calculate the average \fesc of the stack. The result is \Nstack sets of distributions of 10,000 IGM transmission values through the F275W filter, for the \Nstack galaxies in the entire redshift range.
Then for any individual stack consisting of $N_{\rm gal}$ galaxies, we take 
the arithmetic mean of the corresponding $N_{\rm gal}$ independent IGM transmission values from the respective individual transmission distributions, to obtain the mean IGM transmission,
hereafter denoted as $\langle\bar t_{\rm IGM}\rangle$, which shows a single-peaked distribution \citep{bassettIGMTransmissionBias2021}.
In Fig.~\ref{fig:IGMtrans}, we show the mean IGM transmission distributions for all the 10 stacks. We notice that in both redshift bins, the mean IGM transmission for the full galaxy sample has the smallest width, as compared to the corresponding subsamples, since the full sample has the largest number of galaxies in each redshift bin.
We rely on $\langle\bar t_{\rm IGM}\rangle$ to break the degeneracy between $t_{\rm IGM}$ and \fesc, since the distribution of $\langle\bar t_{\rm IGM}\rangle$ is computed from the MC approach, adequate for our large galaxy samples selected from a wide sky coverage of 426 arcmin$^2$ in total.
The median values and the 1-$\sigma$ error bars corresponding to the [16th, 84th] percentiles are given in Table~\ref{tab:fesc_stack}.
We perform bootstrap random draws from the $\langle\bar t_{\rm IGM}\rangle$ distribution to derive the constraints on \fesc, in order to take into account the sample variance of the mean IGM transmission for each stack.

To obtain the appropriate values of the intrinsic flux ratio of UVC to LyC, we run extensive sets of CIGALE SED fitting analysis of all galaxies within the entire stacking samples, following the numerical setup described in Sect.~\ref{subsect:sed}. We do not use the observed F275W flux to retrieve model independent intrinsic SED at the LyC wavelengths of all galaxies, adding together the various components (\ie stellar, nebular lines and continuum, and dust) without IGM and ISM absorption. We then compute the inner products of this intrinsic SED and the filter throughput of F275W and F606W, covering the rest-frame LyC and 1,500 \AA\ UVC, respectively, to estimate the intrinsic flux of LyC and UVC for each galaxy in the stacking samples. Finally, we take a median of all the intrinsic UVC to LyC flux ratios of galaxies that reside in each one of the stacks as the sample average $\left( \frac{F_{\rm UVC}}{F_{\rm LyC}} \right)_{\rm int}$, shown in Table~\ref{tab:fesc_stack}. We note that these intrinsic ratios are consistent with the conventional values quoted in the literature, \citep[\eg][]{Guaita.2017,Rutkowski.2017,Smith.2018,Alavi.2020}.

Then the relative and absolute LyC escape fractions are connected via
\begin{align}\label{eq:fesc_abs}
    \fesc^{\rm abs} = \fesc^{\rm rel} \,10^{-0.4 A_{\rm UV}},
\end{align}
where $A_{\rm UV} = 10.33 \,E(B-V)$ following the Calzetti dust attenuation law \citep{Calzetti.2000} appropriate for high-$z$ star-forming galaxies.
We take the sample median value of E(B-V) as the default dust extinction estimates when computing $\fesc^{\rm abs}$.
We adopt the $L_{\rm UV}$ normalization in stacking described in Eq.~\ref{eq:Luv_wht} in Sect.~\ref{subsect:stack}.
To properly take advantage of our dedicated IGM transmission MC simulations, we perform bootstrap resample to randomly draw 3,000 values from the mean IGM transmission distribution for each stack, in order to incorporate the sample variance in $\langle\bar t_{\rm IGM}\rangle$ in the \fesc estimates.
As a consequence, the measured physical properties for all our stacking galaxy samples are summarized in Table~\ref{tab:fesc_stack}.
We emphasize that the reported isophotal magnitudes (or upper limits) therein are for a galaxy at the respective median redshift and median $M_{\rm UV}$ of the stack.

\begin{figure*}[t!]
    \centering
    \includegraphics[width=.8\textwidth,trim=0cm -0.4cm 0cm 0cm,clip]{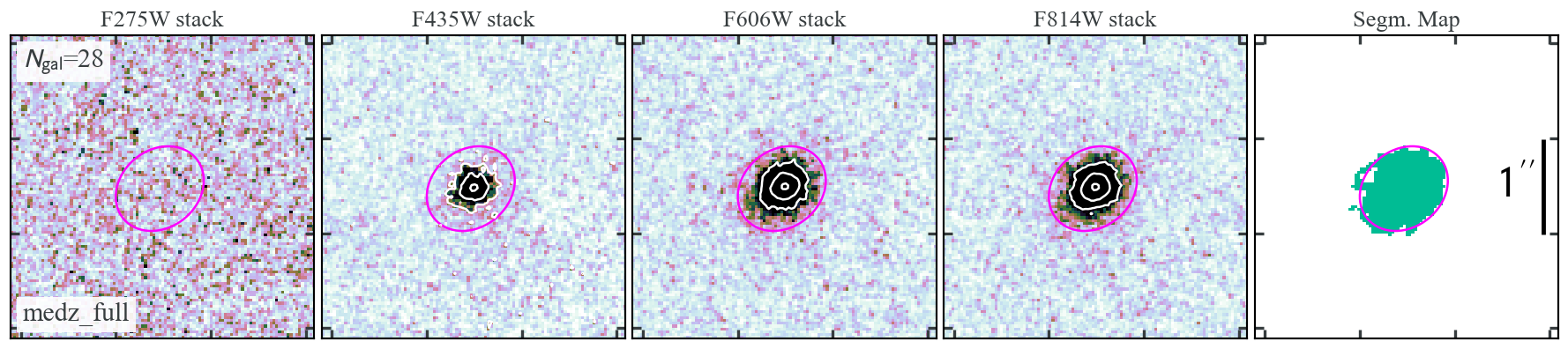}
    \includegraphics[width=.19\textwidth,trim=0.1cm 0cm 0cm 0cm,clip]{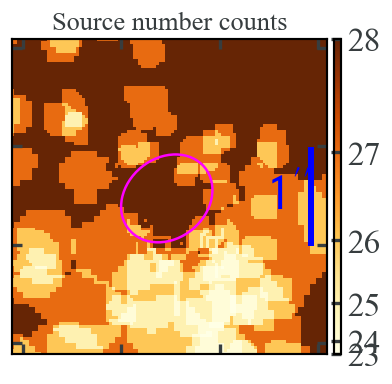}\\
    \vspace*{-.5em}
    \caption{
    \label{fig:stack_cnt}
    The multi-band stacked images and the segmentation map utilized for photometric measurements of the full \medz galaxy sample in the redshift range of $z\in[2.5,3.0]$.
    {\bf (Left five panels)}: the stacked flux stamps in filters F275W, F435W, F606W, and F814W, created using the $L_{\rm UV}$ normalization prescribed in Eq.~\ref{eq:Luv_wht}.
    The fifth panel from the left shows the segmentation region defined using the F606W stack as 
    detection image, where we derive the isophotal flux in each stack (see Sect.~\ref{subsect:photom} for more details).
    All stamps have $3\arcsec\times3\arcsec$ and a plate scale of $0\farcs03$.
    The magenta ellipse overlaid marks the Kron elliptical aperture \citep{kron.1980}, matching very well the F606W isophot.
    The white contours overlaid in the non-LyC filters correspond to the 10, 30, and 90 percentiles of the peak flux in respective filters, to highlight the good performance of our re-centroiding procedure elaborated in Sect.~\ref{subsect:stack}.
    {\bf (Right most panel)}: the number count map showing the number of individual sources contributing to each one of the spatial pixels.
    The number reaches 28 in the center indicating a good re-centroiding process before the stacking. 
    The deficit in the outskirts of the source count map comes from masked nearby neighbors of individual galaxies.
    }
\end{figure*}

We first show the stacked image stamps of the \medz full sample as an example of the end products from our stacking analyses elaborated in Sect.~\ref{sect:stack}.
The combined images stacked using the $L_{\rm UV}$ normalization of all four filters (F275W, F435W, F606W, and F814W) on the entire $3\arcsec\times3\arcsec$ spatial extent with $0\farcs03$ plate scale are shown in Fig.~\ref{fig:stack_cnt}.
We also exhibit the segmentation map defined in F606W via our \textsc{Photutils} photometry with the default parameters. The 10, 30, and 90 percentiles of the peak flux in each non-LyC filter are also marked as white contours. 
We see that our rigorous stacking methodology ensures a centrally concentrated light profile in the three non-LyC filters that probe the rest-frame UV spectra of the sample galaxy population.
The source number counts as shown in the far right panel of Fig.~\ref{fig:stack_cnt} implies a good re-centroiding process during stacking.

\begin{figure*}[h!]
    \centering
    \includegraphics[width=\textwidth,trim=0cm 3.7cm 0cm 3.2cm,clip]{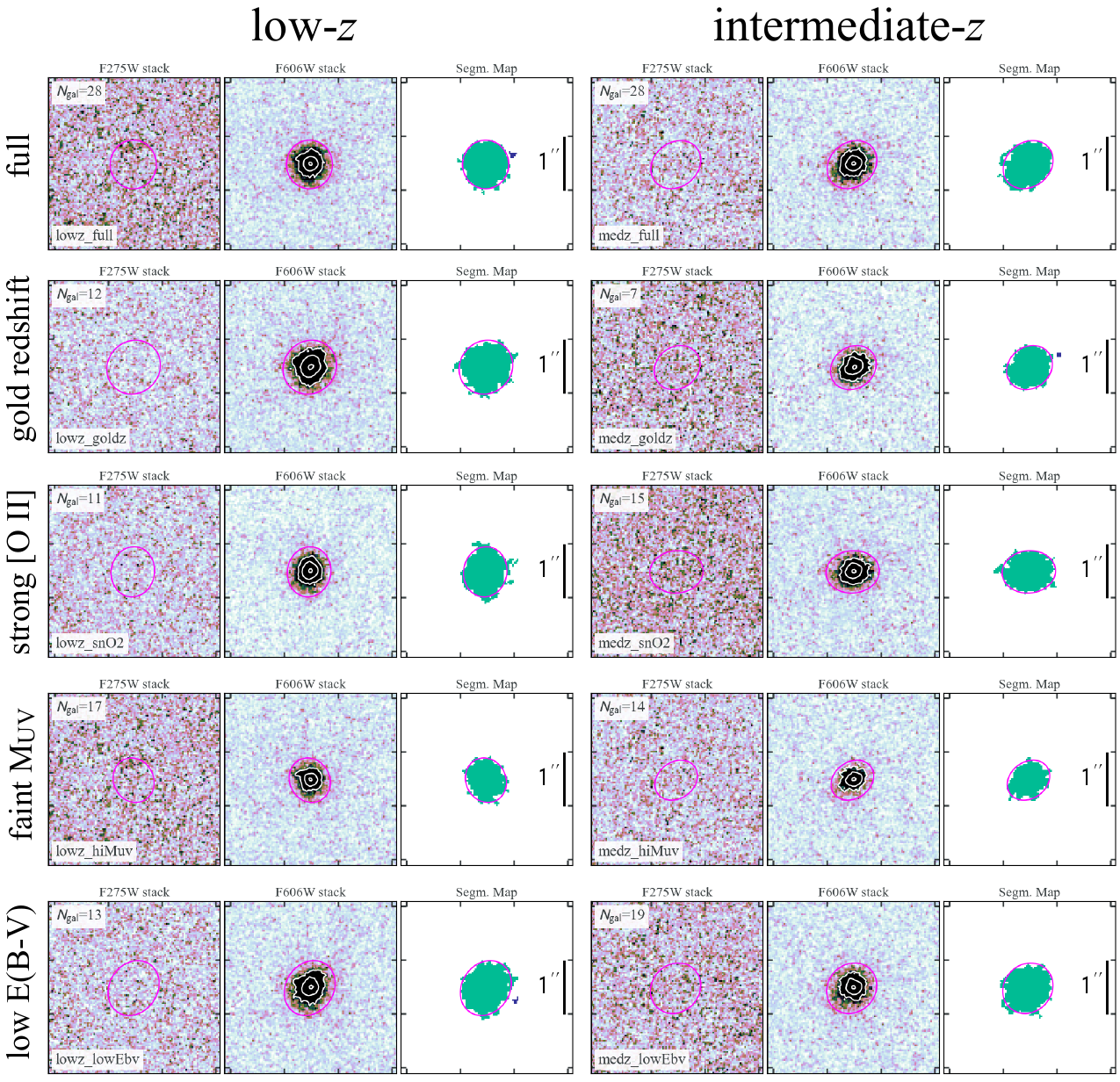}\\
    \vspace*{-.5em}
    \caption{\label{fig:stack_montage}
    LyC and UVC stacks of all the full and sub samples separated into two redshift bins: \lowz (the left three columns) and \medz (the right three columns). From top to bottom, the stacking results are shown for the full sample as well as the \goldz, \snOII, \hiMuv, \lowEbv sub-samples for each redshift bin.
    The fluxes of LyC and UVC are covered in filters F275W and F606W respectively. All stacks are created using the $L_{\rm UV}$ normalization outlined in Eq.~\ref{eq:Luv_wht}.
    As in Fig.~\ref{fig:stack_cnt}, we also show the segmentation map for each stack, where we derive the isophotal fluxes. The magenta ellipse corresponds to the Kron elliptical aperture, using which we perform the empty aperture analysis to derive upper limits on $\left( \frac{F_{\rm LyC}}{F_{\rm UVC}} \right)_{\rm obs}$, since all F275W stacks show non-detections of the escaping LyC signals.
    The white contours overlaid mark the 10, 30, and 90 percentiles of the peak flux for each F606W stack.
    All stamps share the same size of $3\arcsec\times3\arcsec$ with a 30 milli-arcsecond plate scale.
    }
\end{figure*}

Our large parent sample size allows us to constrain \fesc in two separate redshift bins: the \lowz sample of $z\in[2.4,2.5]$ with $z_{\rm median}=2.44$ and the \medz sample of $z\in[2.5,3.0]$ with $z_{\rm median}=2.72$, to account for the rapidly evolving IGM opacity across redshifts.
Although the full samples of both redshift bins have equal sample size (i.e. 28 galaxies each), we caution that the derived \fesc values are not comparable to each other since the F275W filter probes different rest-frame wavelengths.
The stacked image stamps of F275W (covering LyC) and F606W (covering the rest-frame 1500 \AA\ flux as UVC) and the segmentation maps for the full and sub samples of both redshift bins are shown in Fig.~\ref{fig:stack_montage}.

All 10 F275W stacks exhibit non-detection of escaping LyC signals. Consequently, we perform random empty aperture analyses as in \citet{Prichard.2022}, to estimate upper limits on the observed ratio of LyC flux to UVC flux.
For each of the 10 stacking samples, we randomly draw 3,000 Kron elliptical apertures (represented by magenta ellipses in Fig.~\ref{fig:stack_montage}) that closely correspond to the F606W isophotes in the non-detection F275W stacked images. We then compute the sum of the fluxes within these apertures, treating them as individual measurements of the F275W isophotal fluxes.
This practice provides a range of estimates of the observed LyC/UVC flux ratios for each stack. 
Following \citet{gehrelsConfidenceLimitsSmall1986},
we take the 84.13 percentile of this range as the 1-$\sigma$ upper limit on $\left( \frac{F_{\rm LyC}}{F_{\rm UVC}} \right)_{\rm obs}$, reported in Table~\ref{tab:fesc_stack}.
Substituting the 3,000 individual measurements of $\left( \frac{F_{\rm LyC}}{F_{\rm UVC}} \right)_{\rm obs}$ and 3,000 bootstrap random draws from the $\langle\bar t_{\rm IGM}\rangle$ distribution shown in Fig.~\ref{fig:IGMtrans} into Eq.~\ref{eq:fesc_rel} yields a histogram of 3,000 $\fesc^{\rm rel}$ estimates for each stack. The 84.13 percentile of this histogram is adopted as the 1-$\sigma$ upper limit on $\fesc^{\rm rel}$ \citep{gehrelsConfidenceLimitsSmall1986}. Subsequently, the upper limits on $\fesc^{\rm abs}$ are determined using Eq.~\ref{eq:fesc_abs}.
All the upper limits on relative/absolute escape fractions derived from the 10 stacks are reported in Table~\ref{tab:fesc_stack}.

We measure a 1-$\sigma$ lower limit of mag$_{\rm F275W}>28.7$ ABmag and a 1-$\sigma$ upper limit of $\left( \frac{F_{\rm LyC}}{F_{\rm UVC}} \right)_{\rm obs}<0.026$ for the \lowz full sample using the $L_{\rm UV}$ normalization.
Folding in the sample variance of the mean IGM transmission for the \lowz full sample, we derive an upper limit of $\fesc^{\rm rel}<36.8\%$ at 1-$\sigma$ confidence level.
On average, the \lowz sample shows slightly higher dust attenuation of E(B-V)=0.21 than the \medz full sample.
After correcting for dust, we get a stringent 1-$\sigma$ upper limit of $\fesc^{\rm abs}<5.0\%$ from our 28 galaxies spectroscopically confirmed at $z\in[2.4,2.5]$.
We achieve looser 1-$\sigma$ upper limits of $\fesc^{\rm rel}<113.1\%$ and $\fesc^{\rm abs}<26.1\%$ for the \medz full sample, partly due to declining IGM transmission at high redshifts, such that $\langle\bar t_{\rm IGM}\rangle$ has a lower median value of $\sim$20\% at $z\sim2.7$ than $\sim$37\% at $z\sim2.4$, according to our MC line-of-sight IGM attenuation simulations.

Our large sample size enables us to test the strengths of the escaping LyC flux as a function of galaxy global properties.
Within both redshift bins, we further select four subsamples with different criteria defined in Sect.~\ref{subsec:inspect}.
We first repeat our stacking analysis on the \goldz sub samples, comprising galaxies in the first and second redshift quality quintiles ranked by our dedicated visual inspection efforts, considered to have the most secure spectroscopic redshifts.
There are \Glowz and \Gmedz galaxies in the \goldz sub samples in the \lowz and \medz redshift bins, respectively.
We achieve a tight 1-$\sigma$ upper limit on the observed LyC-to-UVC flux ratio of $\left( \frac{F_{\rm LyC}}{F_{\rm UVC}} \right)_{\rm obs}<0.017$ from the \lowz \goldz subsample. This yields stringent upper limits of 
$\fesc^{\rm rel}<23.7\%$ and $\fesc^{\rm abs}<3.7\%$ at 1-$\sigma$ confidence level. The \medz \goldz subsample produces a comparable 1-$\sigma$ upper limit of $\fesc^{\rm abs}<22.5\%$, similar to that from the full \medz sample.

Next, we perform stacking and photometry on the \snOII sub samples. Thanks to the wide coverage of the \hst WFC3/G141 slitless spectroscopy in the \uvc fields, the vast majority of our galaxies have rest-frame near-UV and optical SED measured by G141. The redshift cut imposed by our original sample selection (\ie $z\gtrsim2.4$) precludes our access to some frequently used strong rest-frame optical nebular emission lines, \eg, \Ha, \OIII. The only strong line covered by G141 at this redshift range is \OII. As such, we take advantage of the CHArGE initiative that reanalyzes all the existing \hst NIR imaging and slitless spectroscopy in our fields, to obtain the flux, 1-$\sigma$ uncertainty, equivalent width of \OII for our galaxies, as briefly mentioned in Sect.~\ref{subsec:inspect}.
There are \Olowz and \Omedz galaxies in the \snOII sub samples in the \lowz and \medz redshift bins, respectively.
After incorporating the full distribution of the mean IGM transmission and the observed LyC-to-UVC flux ratio, we obtain the 1-$\sigma$ upper limits of $\fesc^{\rm abs}<4.4\%$ and $\fesc^{\rm abs}<29.0\%$ for the \lowz and \medz \snOII sub samples, respectively.

Furthermore, we also conduct the stacking analysis to the galaxy subgroups with intrinsically faint $M_{\rm UV}$ (the \hiMuv sub samples) and the galaxies with low dust extinction (the \lowEbv sub samples). For the \hiMuv sub sample selection, we require $M_{\rm UV}>-20.5$, comparable to the characteristic $M_{\rm UV}^{\ast}$ of the UV luminosity functions at these redshifts \cite{bouwensGalaxiesMagnifiedHubble2022a}. For the \lowEbv sub sample selection, we require E(B-V)<0.2, since rich dust content is detrimental to the escape of LyC flux.
We indeed obtain looser upper limits on the LyC escape fractions from these two subsamples in both redshift bins than those from their corresponding \goldz subsamples with equal or less number of sources. 
Nevertheless, we notice that all our 10 stacks show LyC flux non-detections.
These non-detections of F275W flux might stem from the CTE degradation effect.
Yet it seems unlikely since the LyC stacks from \citet{Smith.2020} show no evidence of CTE effects when comparing the pristine WFC3/UVIS ERS data \citep{Windhorst.2011} to the later observed HDUV data \citep{Oesch.2018xn}, which is more prone to CTE degradation.
If assuming the median IGM transmission value for our samples, we get consistent 1-$\sigma$ upper limits below $\fesc^{\rm abs}<35\%$ from the 10 samples, generally compatible with some recent findings by \citet{jung.2024} at similar redshifts.

\section{Conclusions}\label{sect:conclu}

In this paper, we introduce the \uvc \hst Cycle-26 Treasury Program (\hst-GO-15647, PI: Teplitz), awarded in total 164 orbits of primary WFC3/F275W and coordinated parallel ACS/F435W imaging observations. Its wide sky coverage of $\sim$426 arcmin$^2$, a factor of 2.7 larger than all previous \hst UV data combined, makes it the largest space-based UV sky survey of distant galaxies with high angular resolution. We present its first set of science-enabling data products --- the coadded imaging mosaics publicly released at \url{https://archive.stsci.edu/hlsp/uvcandels} --- togther with an in-depth description of the data reduction procedures.  These mosaics are astrometrically aligned to the \candels world-coordinate system.  The data processing is designed to overcome the issues affecting the quality of UVIS imaging: the CTE degradation, readout cosmic ray corrections, epoch-varying scattered light during the CVZ observations, etc. As a highlight of the research potential of the UVCANDELS dataset, we focus on one particular application of these imaging mosaics, \ie, searching for strong individual star-forming galaxies that leak LyC radiation and constraining the LyC escape fraction through stacking non-detections.

Our main conclusion is summarized as follows.
\begin{itemize}
    \item We build a large compilation of currently existing spectroscopic data sets in the \candels fields and organize dedicated visual inspection efforts to vet the quality of the rest-frame UV/optical spectra for sources at $2.4\lesssim z\lesssim 3.0$, to exclude spurious spectroscopic redshift measurements.
    \Nstack galaxies with non-detections in F275W (SNR<3) are selected as the parent stacking sample.

    \item We design a rigorous and efficient image stacking methodology using the $L_{\rm UV}$ normalization. 
    Our stacking method is capable of properly recentroiding cutout stamps in terms of source optical/UV isophotes, masking any possible neighboring contaminants, and subtracting local residual sky backgrounds.

    \item We apply our stacking methodology to galaxies in two separate redshift bins: the \lowz ($z\in[2.4,2.5]$) and the \medz ($z\in[2.5,3.0]$) full samples, both consisting of 28 galaxies, respectively.
    We report no significant (>2-$\sigma$) detection of escaping LyC fluxes from both stacks.
    To properly take into account the highly stochastic IGM attenuation of ionizing radiation, we conduct MC line-of-sight IGM transmission simulations and thus derive the full distribution of the absolute LyC escape fraction. Taking into account the resulting full distributions, we measure stringent 1-$\sigma$ upper limit of $\fesc^{\rm abs}\lesssim5\%$ and $\fesc^{\rm abs}\lesssim26\%$ from the full \lowz and \medz stacks respectively.
    
    \item We further select the galaxies in each redshift bin to get four subsamples according to their global properties, to which we re-apply our stacking analysis.
    All the subsample stacks similarly exhibit non-detection.
    The stacks at $z\approx2.44$ show consistent 1-$\sigma$ upper limits below $\fesc^{\rm abs}\lesssim7\%$, whereas the stacks at $z\approx2.72$ show $\fesc^{\rm abs}\lesssim35\%$ at 1-$\sigma$ confidence level consistently,
    after incorporating the full sample variance of mean IGM transmission given by our MC simulations.
        
\end{itemize}

As summarized in Sect.~\ref{sect:uvc}, constraining the escape fraction of the ionizing radiation from galaxies and AGNs at $z\gtrsim2.4$ \citep[see our companion work of][for the stacking analysis of AGNs using similar methodology]{smithLymanContinuumEmission2024a} is merely one of the five major science goals of the \uvc program.
The unique UV/optical dataset produced and publicly released by the \uvc team will offer tremendous opportunities of scientific explorations on various aspects of galaxy formation and evolution. The science-enabling data products will also support the near and mid infrared observations from the \jwst in these legacy extragalactic survey fields.

\begin{acknowledgements}
We would like to thank the anonymous referee for very useful comments that help improve the clarity of this paper.
XW is supported by the National Natural Science Foundation of China (grant 12373009), the CAS Project for Young Scientists in Basic Research Grant No. YSBR-062, the Fundamental Research Funds for the Central Universities, the Xiaomi Young Talents Program, and the science research grant from the China Manned Space Project.
This work is based on observations with the NASA/ESA Hubble Space Telescope obtained at the Space Telescope Science Institute, which is operated by the Association of Universities for Research in Astronomy, Incorporated, under NASA contract NAS5-26555. Support for Program number HST-GO-15647 was provided through a grant from the STScI under NASA contract NAS5-26555.
XW acknowledges work carried out, in part, by IPAC at the California Institute of Technology, and was sponsored by the National Aeronautics and Space Administration.
RAW acknowledges support from NASA JWST Interdisciplinary Scientist grants
NAG5-12460, NNX14AN10G and 80NSSC18K0200 from GSFC. 
We thank A. Barger and N. Reddy for providing the reduced spectra of their Keck observations useful for our visual inspection efforts.
We acknowledge A. Inoue for providing the line-of-sight Monte Carlo IGM transmission code.
\end{acknowledgements}

\software{
\sex \citep{Bertin.1996},
\textsc{CIGALE} \cite{Boquien.2019},
\textsc{Photutils} \citep{Bradley.2022},
\grzl \citep{Brammer.2021},
\adriz \citep{Hack.2021},
\textsc{PyRAF} (Science Software Branch at STScI 2012),
the line-of-sight Monte Carlo IGM transmission code by \citet{Inoue.2014},
\tao \citep{bassettIGMTransmissionBias2021}.
}


\appendix

\section{Example figures used for visual confirmation of spectroscopic redshift}
\label{sec:appendixInspect}
As discussed in Sect.~\ref{subsec:inspect}, we visually confirmed the spectroscopic redshift of the galaxies in our sample since some ground based spectra are prone to noise and limited by seeing conditions. Here, we show an example figure of a galaxy that was ranked into each quintile. The blue curve is the spectrum obtained directly from the archive of the referenced literature source, and the red curve is the same curve convolved with a $\sigma$\,=\,3\,\AA\ wide Gaussian. Several regions of the spectra where emission/absorption lines are expected are shown in greater detail for inspection of the line profiles. The $z_{\rm phot}$ shown is taken from \citet{Skelton.2014}. HST image cutouts filter bands are indicated in their own panels, with the LyC filter distinguished. The original HST mosaics where the image cutouts were extracted from is indicated in blue text below all the image cutouts displayed. The $\chi^2$ image refers to images constructed as described in \citet{Szalay.1999}. Individual ranks are shown for several co-authors, as well as the average, median, and standard deviation of these values. 

\subsection{1st quintile}
\begin{figure*}[h!]
    \centering
    \includegraphics[width=.9\textwidth,trim=0cm 0cm 0cm 0cm,clip]{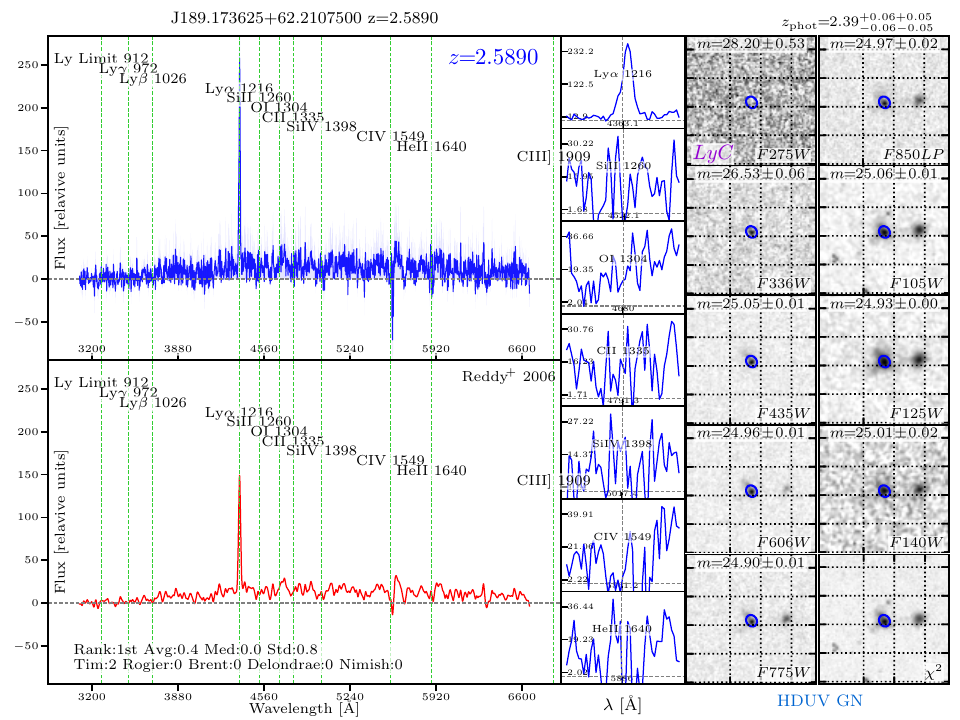}  
\end{figure*}
\clearpage
\subsection{2nd quintile}
\begin{figure*}[h!]
    \centering
    \includegraphics[width=.9\textwidth,trim=0cm 0cm 0cm 0cm,clip]{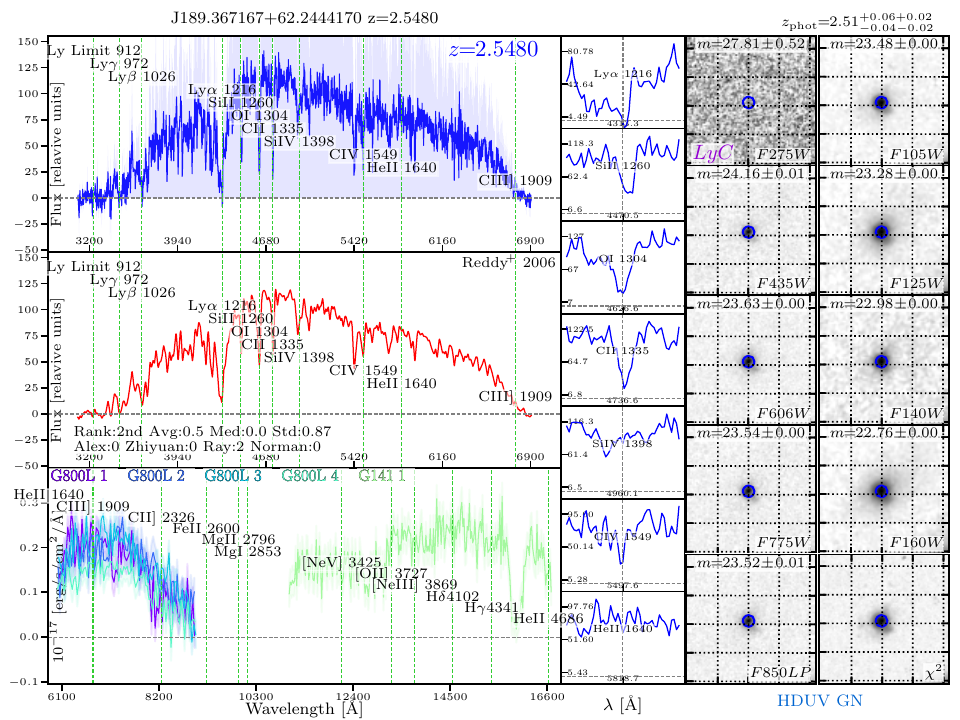}  
\end{figure*}
\clearpage
\subsection{3rd quintile}
\begin{figure*}[h!]
    \centering
    \includegraphics[width=.9\textwidth,trim=0cm 0cm 0cm 0cm,clip]{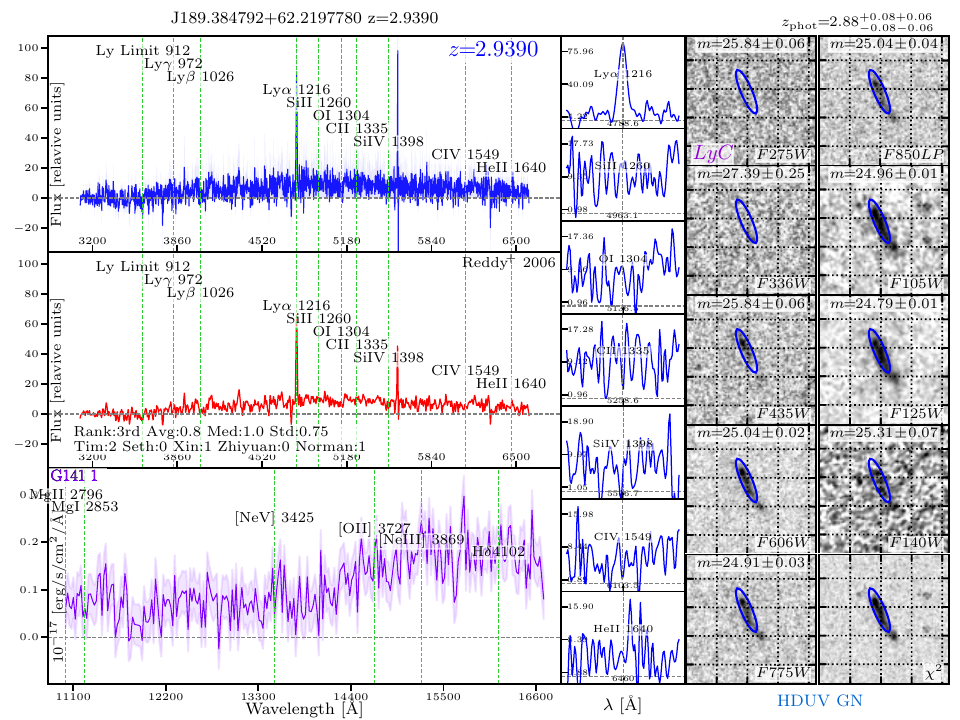}  
\end{figure*}
\clearpage
\subsection{4th quintile}
\begin{figure*}[h!]
    \centering
    \includegraphics[width=.9\textwidth,trim=0cm 0cm 0cm 0cm,clip]{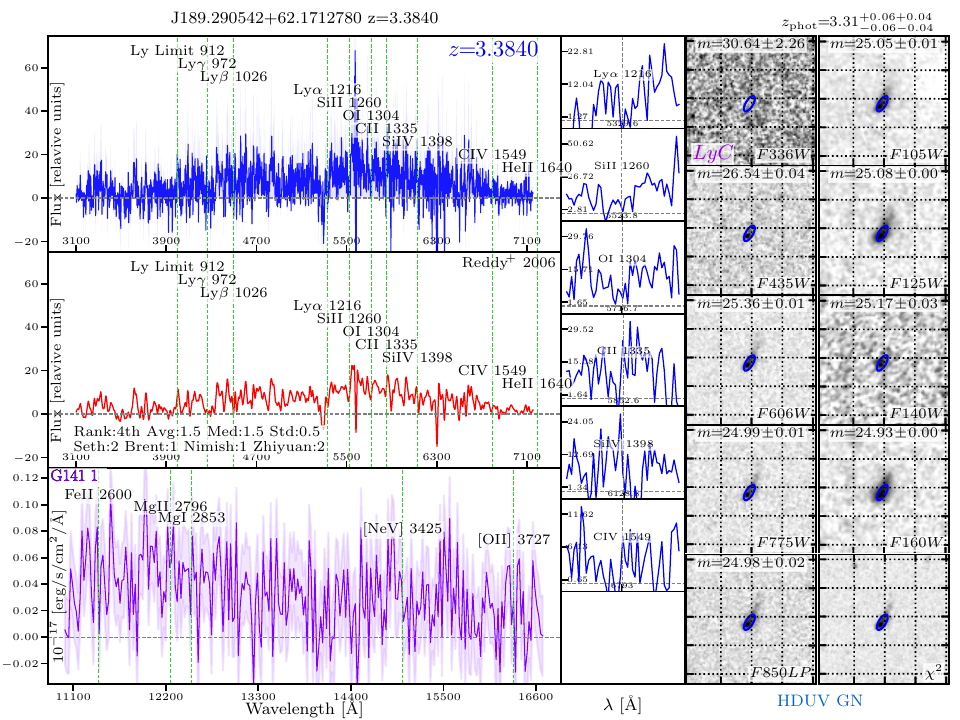}  
\end{figure*}
\clearpage
\subsection{5th quintile}
\begin{figure*}[h!]
    \centering
    \includegraphics[width=.9\textwidth,trim=0cm 0cm 0cm 0cm,clip]{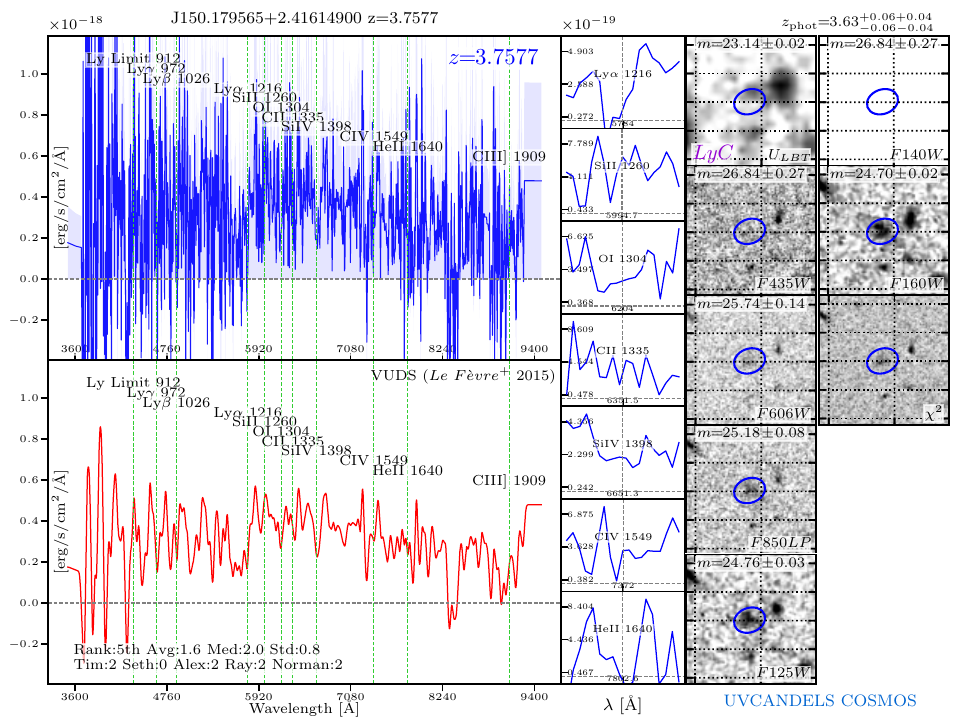}  
\end{figure*}

\clearpage

\section{Individual galaxies with high-quality $z_{\rm spec}\geq2.4$ showing significant F275W detections} 
\label{appendix:indvd_lyc}

{
\tabletypesize{\scriptsize}
\tabcolsep=2pt
\begin{deluxetable*}{cccccccccccccccccc}   \tablecolumns{17}
    \label{tab:indvd_LyC}
    \tablewidth{0pt}
    \tablecaption{
    Properties of the five galaxies showing significant WFC3/F275W flux (SNR$\geq$3) from the \uvc photometric catalog.
    They all have secure spectroscopic redshifts at $z\geq2.4$ confirmed by ground-based and/or \hst grism spectroscopy.
    To ensure that our stacking analysis is not dominated by a few bright sources,
    we exclude them in our stacking analysis, but present their detailed photometric measurements and spectroscopic redshifts here.
    }
\tablehead{
 \colhead{Field} & 
 \colhead{ID} & 
 \colhead{R.A.} & 
 \colhead{Decl.} & 
 \colhead{$z_{\rm spec}$} & 
 \colhead{Ref.\tablenotemark{a}} & 
 \multicolumn{7}{c}{\hst photometry [ABmag]} \\
    & & [deg.] & [deg.] &  &  &  \multicolumn{7}{c}{\hrulefill} \\
    & & & & & & 
 \colhead{F275W} & 
 \colhead{SNR$_{\rm F275W}$} & 
 \colhead{F435W} & 
 \colhead{F606W} & 
 \colhead{F814W} & 
 \colhead{F125W} & 
 \colhead{F160W} 
}
\startdata
  GOODS-N & 00453 & 189.159046 & 62.115474 & 3.656 & 1     &  26.58 & 3.0 & 26.13 & 24.28 & 23.98 & 24.09 & 23.92 \\ 
  GOODS-N & 06226 & 189.179527 & 62.185702 & 3.231 & 1,2,3 &  25.49 & 5.0 & 24.52 & 23.38 & 23.00 & 22.70 & 22.32 \\ 
  GOODS-N & 20748 & 189.378395 & 62.281930 & 2.504 & 3,4   &  26.35 & 3.4 & 25.30 & 24.61 & 24.53 & 23.88 & 23.44 \\ 
  GOODS-N & 22588 & 189.311919 & 62.296512 & 2.486 & 1,4   &  25.63 & 7.2 & 24.70 & 24.23 & 24.17 & 24.08 & 24.10 \\ 
  EGS     & 21708 & 215.040642 & 52.995283 & 2.408 &  3    &  26.51 & 3.1 & 25.60 & 24.97 & 24.82 & 24.31 & 24.05 
\enddata
    \tablenotetext{a}{The references where these spectroscopic redshifts are sourced from. 1: \citet{Barger.2008}. 2: CHArGE \citep{Kokorev.2022}. 3: MOSDEF \citep{Kriek.2015}. 4: \citet{Reddy.2006}.}
\end{deluxetable*}
}

\begin{figure*}[h!]
    \centering
    \includegraphics[width=.79\textwidth,trim=0cm 6.5cm 0cm 1.5cm,clip]{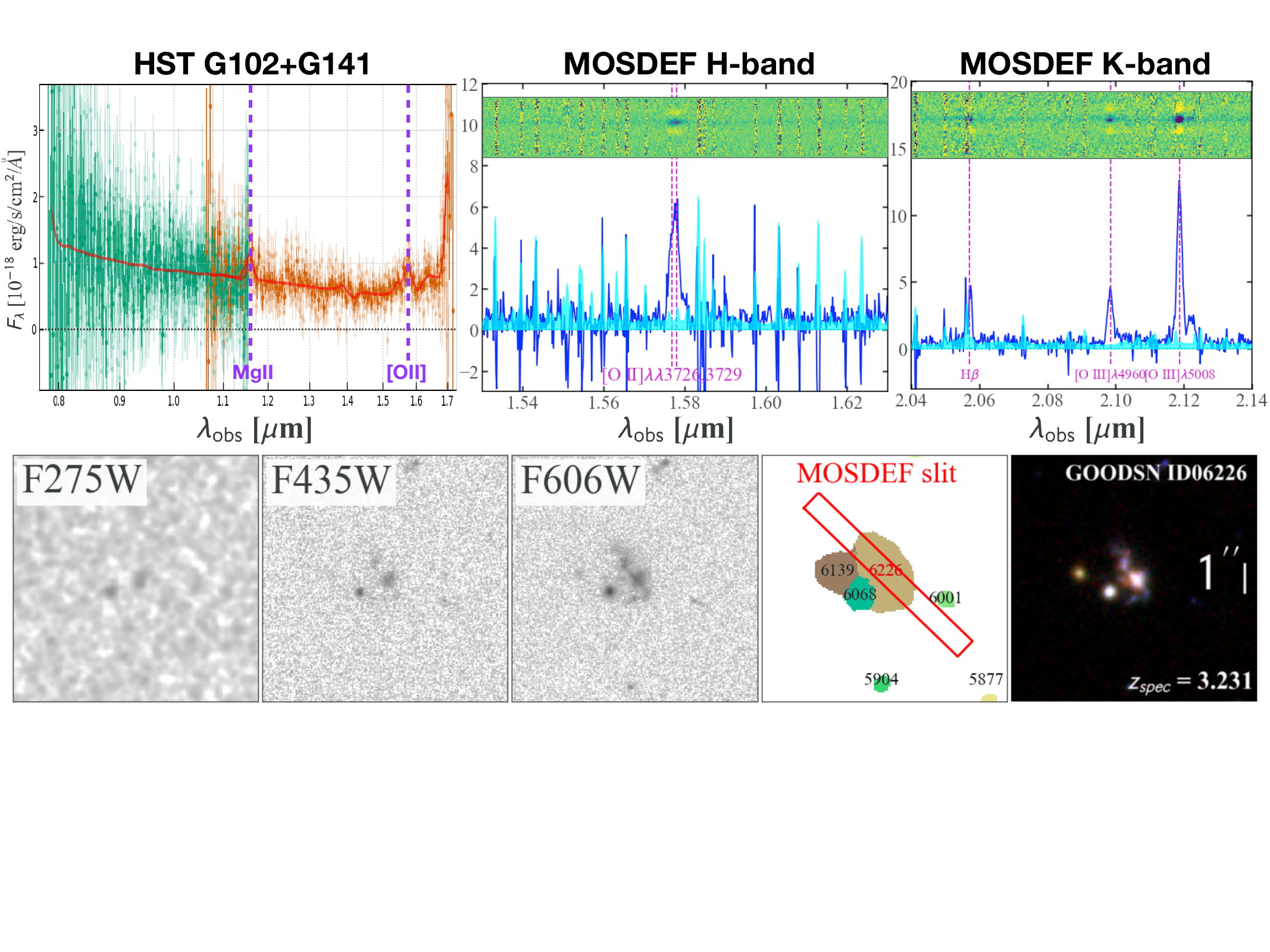}
    \vspace*{-1em}
    \caption{\label{fig:ID06226}
    Galaxy ID 6226 at $z=3.23$ in the GOODS-N field, one example of the five galaxies showing significant F275W detections.
    {\bf Top}: the spectra of ID 6226 taken from \hst G102+G141 slitless spectroscopy and MOSDEF (in H and K bands), showing strong emission features marked by the vertical dashed lines in magenta. The multiple pronounced emission features pinpoint its spectroscopic redshift securely at $z=3.23$.
    The orientation of the MOSDEF slit is shown in the segmentation map displayed in the bottom row.
    {\bf Bottom}: multi-wavelength (F275W, F435W, F606W) postage stamps obtained by \hst, the F160W segmentation map, and the color-composite image produced from \hst imaging. The F275W stamp is smoothed by a boxcar kernel with a width of 0\farcs1 to highlight the detection.
    This galaxy has ${\rm mag_{F275W}}=25.5$ ABmag detected at a high significance of 5$\sigma$.    
    }
\end{figure*}

In total, five galaxies are identified from the catalog of high-quality spectroscopic redshifts within the \uvc survey footprints, which show relatively strong detection in F275W (\eg SNR$\geq$3) and in the redshift range of $z\geq2.4$. To ensure that our stacking analysis is not biased by a few sources with bright fluxes, we exclude them in the stacking analysis. In Table~\ref{tab:indvd_LyC} we present their multi-wavelength photometric measurements for completeness.
We also show the archival spectra taken from ground-based and/or \hst grism instruments \citep{Barger.2008,Kriek.2015,Kokorev.2022} and the HST image stamps of galaxy ID 6226 in Fig.~\ref{fig:ID06226} as an example.
Since GOODS-N is one of the most extensively studied extragalactic fields, there exists a wealth of imaging and spectroscopic data.
In particular, the entire field is covered by the deep Keck/\mosfire infrared spectroscopy (2 hrs each in H and K bands) acquired by the MOSDEF program \citep{Kriek.2015}.
Furthermore, in the whole field these also exists \hst grism spectroscopy of both G102 (2 orbits by \hst-GO-13420, PI: Barro) and G141 (2 orbits by \hst-GO-11600, PI: Weiner).
From these existing deep \mosfire H- and K-band spectroscopy,
we see pronounced nebular emission features of \OIII, \Hb, and \OII lines, with fluxes being $20.7\pm0.3$, $5.4\pm0.7$, and $9.8\pm0.6$ in units of $10^{-17}$~\Funit, respectively.
The \OII and Mg II lines are also clearly detected in \hst grism spectroscopy. 
The wide wavelength coverage (0.8-1.7$\mu$m) of this joint G102-G141 spectroscopy at Hubble's angular resolution basically rules out any possibilities of foreground contamination, since only the spectral features at $z=3.23$ are seen.
ID06226's F275W magnitude is 25.5 ABmag detected at a 5-$\sigma$ significance, and its apparent magnitudes in \B, \V, \I, \J, and \H bands are 24.5, 23.4, 23.0, 22.7, and 22.3 ABmag, respectively.
\clearpage
\bibliography{xinlib,refnew}{}
\bibliographystyle{aasjournal}

\end{document}